\documentclass[preprint,aps,prd,showpacs,superscriptaddress,nofootinbib,tightenlines]{revtex4}

\usepackage{amsfonts}
\usepackage{multirow}
\usepackage{mathrsfs}
\usepackage{graphicx}
\usepackage{amsmath}
\usepackage{amssymb}
\usepackage{bm}
\usepackage{bbm}
\usepackage{color}
\usepackage{slashed}
\usepackage{diagbox}
\usepackage{ulem}
\usepackage{slashbox}
\usepackage{wrapfig}
\usepackage{makecell}

\def\OMIT#1{}

\newcommand{\nn}{\nonumber}

\newcommand{\beq}{\begin{equation}}
\newcommand{\eeq}{\end{equation}}
\newcommand{\bqa}{\begin{eqnarray}}
\newcommand{\eqa}{\end{eqnarray}}
\newcommand {\bseq}{\begin{subequations}}
\newcommand {\eseq}{\end{subequations}}

\begin{document}
\title{\mbox{}\\[11pt]
Near-the-origin divergence of Klein-Gordon wave functions
for hydrogen-like atoms and operator product expansion}


\author{Yingsheng Huang~\footnote{huangys@ihep.ac.cn}}
\affiliation{Institute of High Energy Physics, Chinese Academy of Science, Beijing 100049,
China\vspace{0.2cm}}
\affiliation{School of Physics, University of Chinese Academy of
 Sciences, Beijing 100049, China\vspace{0.2cm}}

\author{Yu Jia~\footnote{jiay@ihep.ac.cn}}
\affiliation{Institute of High Energy Physics, Chinese Academy of Science, Beijing 100049,
China\vspace{0.2cm}}
\affiliation{School of Physics, University of Chinese Academy of
 Sciences, Beijing 100049, China\vspace{0.2cm}}

\author{Rui Yu~\footnote{yurui@ihep.ac.cn}}
\affiliation{Institute of High Energy Physics, Chinese Academy of Science, Beijing 100049,
China\vspace{0.2cm}}
\affiliation{School of Physics, University of Chinese Academy of
 Sciences, Beijing 100049, China\vspace{0.2cm}}

\date{\today}

\begin{abstract}
There have been some long-standing puzzles related to the Coulomb solutions of the
Klein-Gordon and Dirac equations, namely how to understand the physics underlying the weakly divergent
near-the-origin behavior of the $S$-wave wave functions for the hydrogen-like atoms.
Taking the Klein-Gordon wave function as a simpler example,
in this work we demonstrate that, with the aid of the renormalization group equation,
this universal short-distance behavior can be successfully taken into account by
the operator product expansion (OPE) formulated in the nonrelativistic effective field theory (EFT), 
which is tailored for Coulombic atoms. 
The key is to include the relativistic kinetic correction in the EFT. 
Somewhat counterintuitively, these universal near-the-origin logarithmic divergences 
can not be addressed by the OPE set up in the relativistic scalar QED. We conclude that 
the Klein-Gordon wave function at a length scale shorter than the electron Compton wavelength 
may cease to make physical significance.
\end{abstract}

\maketitle

\section{Introduction}

Back to the late 1920s, the Klein-Gordon (KG) and Dirac equations in an external electrostatic Coulomb
field, has played a vital role in accounting for the by-then measured hydrogen spectroscopy, 
and more importantly, establishing quantum mechanics as the true science~\cite{Weinberg:1995mt}. 
As everyone knows, KG equation quickly lost its battle with the Dirac equation in predicting 
the fine structure of the hydrogen atom, since the former only describes the relativistic wave equation 
for a fictitious spinless electron. Later, KG equation has resurrected as a viable field equation in 
the relativistic quantum field theory.

Since the birth of relativistic quantum mechanics, it was known that both KG and Dirac wave functions 
for the $S$-wave hydrogen-like atoms are divergent near the origin, albeit very weakly, 
when the electron coincides with the nucleus~\cite{Weinberg:1995mt,Greiner2000}. For the long time, this symptom has been thought of only 
academic interest. The logarithmic singularity would bring pronounced effect only when the electron is extremely close
to the origin. At practical level, the realistic nucleus has a finite size, and 
this distance scale is already many orders-of-magnitude smaller than the radius of the nucleus. 
Even for an idealized nucleus, {\it i.e.}, an infinitely heavy positively-charged point particle, 
the divergence is so weak so the KG and Dirac wave functions
are still square integrable near the origin, thereby does not ruin the canonic probabilistic interpretation 
of wave function.

However, we take a different attitude in this work. Our goal is to unravel the physical origin 
underlying the logarithmically divergent near-the-origin behavior for the relativistic hydrogen wave function.
For simplicity, in this work we will focus on the Coulomb solution of the Klein-Gordon wave function, 
which seems considerably simpler than the Dirac equation, yet still captures some essential piece of physics.
The analogous analysis for the Dirac wave functions for the $nS_{1/2}$ hydrogen atom will be presented 
elsewhere~\cite{Huang:2019:Dirac}.
 
In Ref.~\cite{Huang:2018yyf}, we have showed that, within the framework of the 
nonrelativistic Coulomb-Schr\"{o}dinger effective field theory (EFT), 
the operator product expansion (OPE)~\cite{Wilson:1969zs} technique can be fruitfully applied to decipher the universal short-distance
behavior of the atomic Schr\"{o}dinger wave functions, which are always regular near the origin.
In this work, we extend \cite{Huang:2018yyf} by incorporating the relativistic kinetic correction into 
the Coulomb-Schr\"{o}dinger EFT. We will illustrate that, by conducting the Wilson expansion of 
the product of the nonrelativistic electron field and the HQET-like nucleus field, 
we are able to interpret the universal logarithmic near-the-origin divergence of
the KG wave functions as the Wilson coefficients. Moreover, with the help of the renormalization group equation (RGE), 
one can successfully reproduce the anomalous scaling behavior exhibited in the KG wave function at small $r$.

The motif of this work is also to clarify the nature of the Klein-Gordon wave function.
The Klein-Gordon equation in an external Coulomb field is often viewed as the relativistic wave equation, 
and one is allowed to probe the arbitrarily small distance profile of the wave function. 
As we will see, our OPE analysis based on the relativistic scalar QED would falsify this misconception.
In our opinion, the KG equation actually secretely describes the nonrelativistic rather than
relativistic bound-state wave function. We believe that it does not make too much sense to talk about
the near-the-origin behavior of the KG wave function when $r$ is much smaller than the electron Compton wavelength. 
On the contrary,  when probing the short-range behavior of the KG wave function,
the minimum value of $r$ should be frozen around the Compton length of the electron.
 
The rest of the paper is organized as follows. In section \ref{Recap:Coulomb:KG:equation}, we revisit the Coulomb solution of the KG equation, focus on the weak singularity associated with the $S$-wave KG wave functions near the origin.
In section~\ref{Quantum:perturb:corr:wave:func:origin}, we perform a nonrelativistic reduction to KG equation, then
apply quantum-mechanical perturbation theory to compute the first-order correction to the wave function at the origin.
In section~\ref{sQED:OPE:failure}, we study the local composite operator renormalization and OPE
within a relativistic field theory that contains the scalar QED and heavy nucleus effective theory.
Unfortunately, this theory fail to reproduce the intended logarithmic divergence of the KG wave function near the origin.

In section~\ref{Proper:NREFT}, we switch the gear and set up a proper nonrelativistic EFT to describe the atoms
bounded by the slowly moving spin-0 electrons and a static nucleus through Coulomb interaction.
We explicitly match the UV theory onto the nonrelativistic EFT through one-loop order, and
verify no contact interaction can arise at least at order-$Z^2\alpha^2$.
In section~\ref{Renorm:local:op:varphiN}, we study the renormalization of the
local composite $S$-wave operator composed of the electron and nucleus fields
in this nonrelativstic EFT. It is found that this operator becomes logarithmic divergent
at order $Z^2\alpha^2$, and a multiplicative renormalization is required to render it finite.

Section~\ref{sec:OPE:NREFT} is the main part of the paper.
We establish the Wilson expansion for the product of nonrelativistic
electron field and nucleus field in both momentum and coordinate space,
through order-$Z^2\alpha^2$.
We use the renormalization group equation for the Wilson coefficient
to resum the leading logarithms
to all orders.
Finally we summarize in \ref{sec:summary}.

\section{Coulomb solution of Klein-Gordon equation, $S$-wave radial wave functions near the origin}
\label{Recap:Coulomb:KG:equation}

For the sake of completeness, in this section we briefly
recapitulate the celebrated Coulomb solution of the Klein-Gordon equation,
with primary interest in the near-origin behavior of the KG radial wave function for
hydrogen-like atoms.
In particular, we will concentrate on the coalescence behavior of the $S$-wave wave functions.
For a spinless electron orbiting around an infinitely heavy nucleus carrying
electric charge $Ze$, the relativistic wave equation reads
\beq
\left[\left(E+{Z e^2\over 4\pi r}\right)^2+ \hbar^2 c^2 {\bf \nabla}^2- m^2 c^4\right] \Psi({\bf r})=0.
\label{KG:Equation}
\eeq
As usual, the KG wave function can be separated into the radial and angular part,
\beq
\Psi_{nlm}({\rm r})= R_{nl}(r) Y_{lm}(\hat{\bf r}),
\eeq
where $Y_{lm}$ is the spherical harmonics, with $l$ and $m$ denoting the
orbital angular momentum and the magnetic quantum number.

The radial wave function obeys the following differential equation~\cite{Greiner2000}:
\beq
\left[ {d^2 \over d r^2} + {Z^2\alpha^2-l(l+1)\over r^2} +
{2Z\alpha\,E  \over \hbar c r}+ {E^2-m^2c^4\over \hbar^2 c^2 }\right]
r R_{nl}(r)=0,
\label{Radial:KG:equation}
\eeq
with the fine structure constant $\alpha= {e^2\over 4\pi \hbar c}$.

The bound-state solution is well-known, which yields the corresponding discrete
eigen-energy:
\bqa
E_{nl} &= &  mc^2 \left\{ 1+ { Z^2\alpha^2 \over \left(n-l-{1\over 2} +
\sqrt{\left(l+{1\over 2}\right)^2-Z^2\alpha^2} \right)^2 }
\right\}^{-{1\over 2}}
\nn\\
&= &  mc^2\left\{ 1-{Z^2\alpha^2 \over 2n^2}- {Z^4\alpha^4\over 2n^4}\left({n\over l+{1\over 2}}-{3\over 4}\right)
+\cdots\right\},
\label{Eigen:energy:KG:hydrogen}
\eqa
where the principal quantum number $n=1,2,\cdots$,
and the orbital angular momentum $l=0,\cdots,n-1$ are
non-negative integers.
In the second line of \eqref{Eigen:energy:KG:hydrogen}, we have expanded the
exact energy level in power series of $Z\alpha$~\footnote{It is historically
well known that, the $Z^4\alpha^4$ term yields the wrong fine structure of hydrogen spectroscopy,
thereby the KG equation, which governs the dynamics of a spin-0 electron,
must be abandoned as being the realistic relativistic wave equation to describe
the hydrogen atom~\cite{Weinberg:1995mt}.}.

The solution for radial wave function is~\cite{Greiner2000}
\beq
R_{nl}(r)= N_{nl} \rho^{l^\prime} \, e^{-{\rho\over 2}}\,{}_1F_1(l^\prime+1-\lambda,2l^\prime+2;\rho),
\label{Radial:WF:expression}
\eeq
where ${}_1F_1$ stands for the confluent hypergeometric function, and
\beq
l^\prime \equiv
-{1\over 2}+\sqrt{\left(l+{1\over 2}\right)^2-Z^2\alpha^2},\quad \beta\equiv {2\over \hbar c}\sqrt{m^2c^4-E_{nl}^2},
\quad \rho \equiv \beta r,\quad  \lambda={2 Z\alpha E_{nl}\over \hbar c \beta}.
\label{Shorthands:symbol}
\eeq
In the $Z\alpha\ll 1$ limit, $\beta \approx {2 mc Z\alpha\over \hbar n}= 2/(n a_0)$,
proportional to the reciprocal of the Bohr radius $a_0 \equiv \hbar/(m c Z\alpha)$,
and $\lambda\approx n$.
The normalization constant $N_{nl}$ can be determined by enforcing
the spatial integral of the density of the conserved charged current
in KG equation to be unity,
\beq
\int\!\! dr\, r^2\, R_{nl}^2(r) \left({E_{nl}\over m c^2}  +{Z e^2 \over 4\pi r m c^2}\right)=1.
\label{Normalization:condition}
\eeq
In the $Z\alpha\ll 1$ limit, this constraint is very close to the conventional
normalization condition $\int\!\!dr\,r^2 R^2(r) =1$ for $Z\alpha\ll 1$.

The power-law scaling of the wave function near the origin
can be simply inferred from the $r\to 0$ behavior of \eqref{Radial:KG:equation}:
\beq
R^{\prime\prime}_{nl}  +{2\over r} R^\prime_{nl}  +{Z^2\alpha^2 -l(l+1)\over r^2}R_{nl}=0.
\eeq
Substituting the ansatz $R_{nl}(r)\propto r^{l^\prime}$, one can solve $l^\prime$
from the following quadratic equation:
\beq
l^\prime (l^\prime + 1) = l(l+1)-Z^2\alpha^2,
\eeq
and take the greater root as specified in \eqref{Shorthands:symbol}.
Note the scaling behavior is called ``anomalous'', since it qualitatively differs from that
affiliated with the Schr\"{o}dinger radial wave functions $R^{\rm Sch}_{nl}(r)\propto r^l$.

\begin{figure}[tb]
\centering
\includegraphics[width=0.8\textwidth]{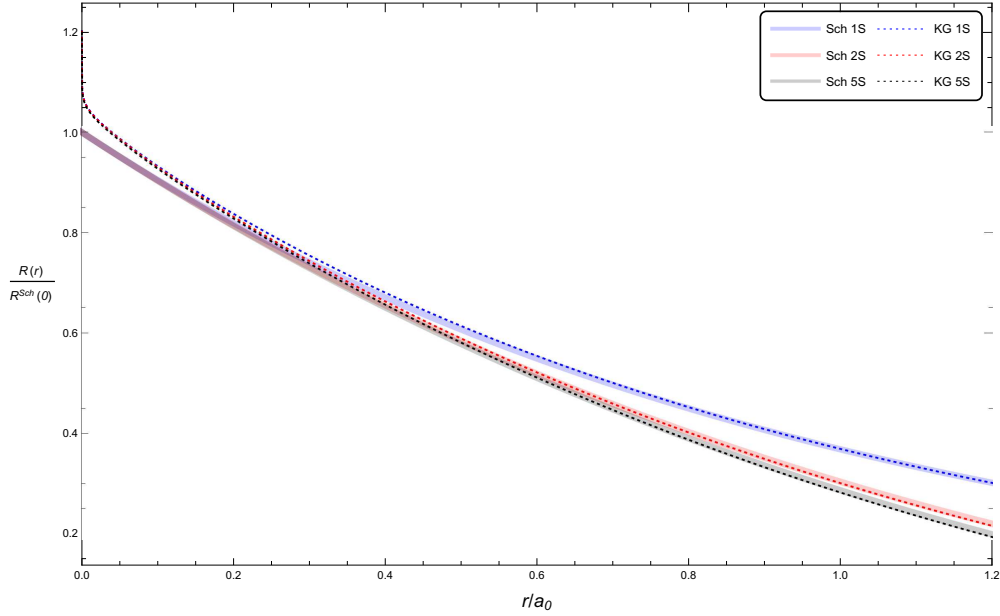}
\caption{The Klein-Gordon and Schr\"{o}dinger radial wave functions for the $1S$, $2S$ and $5S$
hydrogen-like atom. To clearly discern the difference between the KG and Schr\"{o}dinger wave functions,
we deliberately take $Z\alpha={1\over 10}$ with $m=1$.
\label{Fig:Radial:wave:function}}
\end{figure}

In the $Z\alpha\ll 1$ limit, the difference between $l$ and $l^\prime$ is tiny for high $l$.
Nevertheless, for $S$-wave hydrogen-like atom, $l^\prime\approx -Z^2\alpha^2$,
is a very small negative number.
Consequently, the corresponding KG radial wave function at the origin then becomes divergent,
despite being a very weak singularity.

Near the origin, the KG wave function can be approximated as
\bqa
R^{\rm KG}_{n0}(r) &\approx & R_{n0}^{\rm Sch}(0) \rho^{\sqrt{\frac{1}{4}-Z^2\alpha^2}-\frac{1}{2}} \approx
R_{n0}^{\rm Sch}(0)\rho^{-Z^2\alpha^2}
\nn\\
& = & R_{n0}^{\rm Sch}(0)\left[1-{Z^2\alpha^2}\ln{ 2r\over n a_0}+{\mathcal O}
\left(Z^4\alpha^4\ln^2 r\right)\right],
\label{KG:WF:origin:Log:Div}
\eqa
where the expansion is in power series of $Z\alpha$.
Therefore, the wave function near the origin only develops a logarithmic divergence.
It is very important to note that this divergence is {\it universal}, in the sense it applies to
all the $S$-wave Coulombic state, irrespective of $n$ being a discrete or continuum quantum number.

The universal near-the-origin behavior of
the Schr\"{o}dinger wave function is also widely known~\cite{Loewdin1954,Huang:2018yyf}.
For example, for $S$-wave hydrogen-like
atom, one has $R^{\rm Sch}_{n0}(r)= R_{n0}^{\rm Sch}(0)(1-r/a_0)$.
Carefully expanding the $S$-wave KG wave functions from \eqref{Radial:WF:expression},
we find their near-the-origin behaviors to be
\bqa
R^{\rm KG}_{n0}(r) &\approx&   R_{n0}^{\rm Sch}(0)(1- m Z \alpha r) \rho^{-Z^2\alpha^2}
\approx R_{n0}^{\rm Sch}(0)(1-mZ\alpha r)\left(1-Z^2\alpha^2 \ln r +\cdots \right)
\nn\\
& = &  R_{n0}^{\rm Sch}(0)\left(1- m Z\alpha r- Z^2\alpha^2 \ln r + m Z^3\alpha^3 r \ln r + \cdots \right),
\label{KG:Universal:Small:r}
\eqa
where only the {\it universal} terms are exhibited.
For the sake of clarity, in Fig.~\ref{Fig:Radial:wave:function} we plot the various lowest-lying
$S$-wave KG and Schr\"{o}dinger radial wave functions normalized with respect to the corresponding Schr\"{o}dinger
radial wave functions at the origin, from which the universal coalescence behavior encoded in
\eqref{KG:Universal:Small:r} can be seen.

The very goal of this work is to understand the physics underlying these universal short-range terms
from the perspective of quantum effective field theory and renormalization program.

We summarize the key aspects observed for the $S$-wave KG wave functions near the origin,
and raise some questions that will be answered in the following sections:
\begin{itemize}
\item{As indicated in \eqref{KG:WF:origin:Log:Div}, the $S$-wave KG wave functions
exhibit weak logarithmic singularity as $r\to 0$.
This seems to be reminiscent of UV divergence and calls for renormalization.}
\item{The logarithmic coalescence divergence is universal.  That is,
it does not depend on the principle quantum number $n$, regardlessly of being a discrete or continuum label.
Since the wave functions may be regarded as the product of quantum fields sandwiched between the vaccum and the atom,
it is conceivable that their short-range behaviors might be probed with the aid of OPE.
The universal yet regular coalescence behavior observed for Schr\"{o}dinger wave functions,
$R^{\rm Sch}_{n0}(r)= R_{n0}^{\rm Sch}(0)(1-r/a_0)$, has indeed been deciphered by invoking the OPE technique
in the nonrelativistic Coulomb-Schr\"{o}dinger EFT~\cite{Huang:2018yyf}.
Is it possible to further interpret the universal logarithmic divergences as Wilson coefficients related to the OPE?
A crucial difference is that in the Schr\"{o}dinger case, the only nontrivial
Wilson coefficients is of order $Z\alpha$, but
in the KG case, the first nontrivial Wilson coefficient appears to begin at order $Z^2\alpha^2$.}
\item{If the OPE turns out to be the viable arsenal to unravel this riddle,
then what is the appropriate field theory to formulate the product of two fields?
Is it appropriate to start from a relativistic quantum field theory such as the scalar QED,
which is closely tied with the Klein-Gordon equation in an external electrostatic field?}
\item{If the OPE is formulated properly, is it feasible to interpret all the universal terms in
\eqref{KG:Universal:Small:r}} as the Wilson coefficients? That is, can we treat
both regular and divergent Wilson coefficients on an equal footing?

\end{itemize}

\section{Nonrelativistic reduction of KG equation, first-order quantum-mechanical perturbation theory}
\label{Quantum:perturb:corr:wave:func:origin}

Before launching into the field-theoretical formalism, we would like to dig
more physics out of the single-electron wave equation \eqref{KG:Equation}.
Since the spinless electron in a KG hydrogen atom is nonrelativistic, $v/c\sim Z\alpha \ll 1$,
it is useful to carry out a nonrelativistic reduction to the KG equation,
to obtain an effective Schr\"{o}dinger equation.
This reduction procedure is explained in great details in many nice
textbooks~\cite{Holstein2014}, and the resulting effective Schr\"{o}dinger equation reads
\beq
H_{\rm eff} \psi({\bf r})= E\psi({\bf r}),
\eeq
with the effective Hamiltonian
\bseq
\bqa
&& H_{\rm eff} = H_0 + \Delta H= H_0+ H_{\rm kin}+H_{\rm Darwin},
\\
&&  H_0=-\frac{\bm{\nabla}^2}{2m}-\frac{Ze^2}{ 4\pi r},
\label{H0:Coulomb}
\\
&& H_{\rm kin}=-{{\bm\nabla}^4 \over 8 m^3 c^2},\qquad
H_{\rm Darwin}= {1\over 32m^4 c^4}\left[{\bm\nabla}^2,
\left[ \frac{Ze^2}{4\pi r},{\bm \nabla}^2\right]\right].
\label{DeltaH:perturb}
\eqa
\label{Effective:Schr:Hamiltonian}
\eseq
Eq.~\eqref{H0:Coulomb} is the standard nonrelativistic Coulomb Hamiltonian.
Eq.~\eqref{DeltaH:perturb} represents some perturbations due to relativistic effects.
$H_{\rm kin}$ stands for the leading relativistic correction to the kinetic energy,
and $H_{\rm Darwin}$ stands for the Darwin term, which is responsible for the zitterbewegung
motion of the electron~\cite{Holstein2014}.
For simplicity, we have set $\hbar=1$, but still keep $c$ manifest at this stage.
It is seen that $H_{\rm kin}$ makes a contribution of relative order $v^2/c^2$, much more important than the
Darwin term, whose contribution is of relative order $v^4/c^4$.

The first-order corrections to the hydrogen energy level from $\Delta H$ can be found in
virtually every textbook on advanced quantum mechanics (for example, see~\cite{Holstein2014}),
\bseq
\bqa
&& \Delta E^{(1)}_{nl}\Big|_{\rm kin} = \langle nl \vert H_{\rm kin} \vert nl \rangle =
- mc^2 {Z^4\alpha^4\over 2n^4} \left({n\over l+{1\over 2}}-{3\over 4}\right),
\\
&& \Delta E^{(1)}_{nl}\Big|_{\rm Darwin} = \langle nl \vert H_{\rm Darwin} \vert nl \rangle = 0,
\eqa
\eseq
which just reproduces the fine structure of hydrogen,
derived from the exact solutions of the KG equation \eqref{Eigen:energy:KG:hydrogen}.

The structure revealed in \eqref{KG:WF:origin:Log:Div} may suggests us to explore the effect of
first-order correction brought by $\Delta H$ to the Schr\"{o}dinger wave function at small $r$.
For definiteness, let us investigate the perturbative correction to the Schr\"{o}dinger wave function
at the origin for the ground state of hydrogen atom, $\Delta R^{\rm Sch}_{10}(0)$.
In accordance with the standard first-order perturbation theory in quantum mechanics,
we have
\beq
\Delta R^{(1)}_{10}(0)= \sum_{n=2}^{\infty}  R^{\rm Sch}_{n0}(0) {\langle n0 \vert \Delta H \vert 10\rangle \over E_{10}-E_{n0}}
  + \int_0^{\infty}\! {dk\over 2\pi}\, R^{\rm Sch}_{k0}(0) {\langle k0\vert \Delta H\vert 10\rangle \over E_{10}-E_{k0}},
\label{QM:pert:corr:wf:at:origin}
\eeq
where both discrete and continuum intermediate states are included.
Since $\Delta H$ are rotation scalar,
we only need retain those intermediates states with $l=0$.

We are only interested in whether the correction to the wave function at the origin
can lead to divergence or not. For this purpose, we ignore the sum over those
discrete states, which renders finite results.
We just focus on the the high-frequency part of the integration over the continuum $S$-wave states.
The corresponding continuum Coulomb wave functions read~\cite{Landau:1958}:
\bseq
\bqa
&& R_{k0}^{\rm Sch}(r) = \sqrt{8\pi m  Z\alpha k \over 1-e^{-2\pi m Z\alpha\over k}} e^{-ikr}
{}_1F_1\left(1+{im Z\alpha\over k},2;2ikr\right),\qquad E_{k0}={k^2\over 2 m},
\\
 && \int\!\!dr\,r^2 R^{\rm Sch}_{k0} (r) R_{k^\prime 0}^{\rm Sch}(r)=2\pi\delta(k-k').
\eqa
\eseq
where the ``$k/2\pi$-scale'' convention is chosen for normalization~\cite{Landau:1958}.
For brevity, from now on we will adopt the natural unit $\hbar=c=1$.

It is straightforward to compute the matrix elements  encountered in \eqref{QM:pert:corr:wf:at:origin}:
\bseq
\bqa
&& \langle k0\vert H_{\rm kin}\vert 10\rangle = - 2Z^3 \alpha^3
\sqrt{\pi  k \over 2\left( 1-e^{-{2 \pi  m Z\alpha\over k}}\right)}
\left[1-{2 m^2 Z^2\alpha ^2  \cosh {\pi m Z \alpha\over k}
\exp\left({2 m Z\alpha \over k} \tan^{-1}{ m Z\alpha \over k}\right) \over k^2+ m^2 Z^2\alpha^2}\right]
\nn \\
&&  \xrightarrow {k\gg mZ\alpha} -\sqrt{mZ\alpha} Z^3\alpha^3 \left({k\over mZ\alpha}+\frac{\pi }{2}+\cdots \right),
\\
 && \langle k0\vert H_{\rm Darwin}\vert 10\rangle  = -{Z^3\alpha ^3 \over 4 }
 \sqrt{\frac{\pi  k }{2(1-e^{-\frac{2 \pi  mZ\alpha}{k}})}} \left( {k^2\over m^2}+Z^2\alpha ^2  \right)
\\
&& \xrightarrow{k\gg mZ\alpha} -\sqrt{mZ\alpha}\frac{Z^3\alpha^3}{8m^2} \left(\frac{k^3}{mZ\alpha}+\frac{\pi k^2}{2}+\frac{(24+\pi^2)k(mZ\alpha)}{24}+\frac{\pi(\pi^2-24)(mZ\alpha)^{2}}{48}+\cdots \right),
\nn
\label{continuum:matrix:elements}
\eqa
\eseq
where we also presented the expanded expressions for these matrix elements  in inverse powers of $k$,
up to the $k^0$ terms~\footnote{Recall for large wavevector $k$ in \eqref{QM:pert:corr:wf:at:origin},
$R_{k0}(0)\propto k$, and the energy denominator yields a contribution $\propto k^{-2}$.
Thereby it suffice for us to truncate the power series in \eqref{continuum:matrix:elements}
at the order $k^0$, which just generates a mild logarithmic divergence.}.

Simple power counting indicates that the integrations in \eqref{QM:pert:corr:wf:at:origin} are severely UV divergent,
which indicates that the corrections are dominated by contributions from the high-energy intermediate states.
It is natural to imposing a momentum cutoff $\Lambda < m$ in the integral in \eqref{QM:pert:corr:wf:at:origin},
to exclude those dangerous high-energy states that defy a trustworthy description by
nonrelativistic quantum mechanics.  Accordingly,
both $H_{\rm kin}$ and $H_{\rm Darwin}$ bring the following divergent
corrections to the $1S$ wave function:
\bseq
\bqa
\Delta R^{(1)}_{10}(0)\Big|_{\rm kin} &=& R^{\rm Sch}_{10}(0)\left( {Z\alpha\Lambda\over \pi m}+
Z^2\alpha^2 \ln \Lambda+{\rm finite}\right),
\label{Kin:corr:wf}
\\
\Delta R^{(1)}_{10}(0)\Big|_{\rm Darwin} &= & R^{\rm Sch}_{10}(0)
\left({Z\alpha\Lambda ^3\over 24 \pi m^3}+    {Z^2\alpha^2 \Lambda ^2 \over 16 m^2}+ {Z^3\alpha^3\pi  \Lambda\over 24 m}+{\rm finite}\right).
\label{Darwin:corr:wf}
\eqa
\label{UV:Div:Corrs:to:WF:DeltaH}
\eseq
Setting $\Lambda\sim mZ\alpha$, we indeed verify that the kinematic correction term renders a
correction of relative order $v^2\sim Z^2\alpha^2$, and the Darwin term yields a correction $\propto v^4$,
in conformity with the velocity counting rule in \eqref{Effective:Schr:Hamiltonian}.
Although both types of the corrections start from a leading power divergence,
it is important to note that the former also develops a subleading logarithmic divergence,
whereas the latter does not.
Comparing \eqref{Kin:corr:wf} with \eqref{KG:WF:origin:Log:Div}, we find that,
provided that $r$ is replaced with $1/\Lambda$, the coefficients of the logarithmic divergences in both expressions
exactly coincide~\footnote{In the course of the integration over the intermediate states,
if we were replacing the continuum Coulomb states by the free continuum spherical waves,
we might anticipate that the coefficients of the UV divergences would not change.
Unfortunately, this naive anticipation is true only for the {\it leading} UV
divergence, but not for the subleading ones. For example, the coefficient of the logarithmic divergence in
\eqref{Kin:corr:wf} would be only half of the correct value.}.
This is an encouraging sign that the kinetic term $H_{\rm kin}$ might well be the agent
governing the mild logarithmic divergence of the KG wave function near the origin.

The power divergences in \eqref{UV:Div:Corrs:to:WF:DeltaH} are clearly absent in the original KG wave function
\eqref{KG:WF:origin:Log:Div}.
It is suggestive to attribute these severe UV power divergences to our use of
sharp momentum cutoff as the UV regulator, which is known to break our cherished spacetime symmetries.
Thus, these power UV divergences need not to be endowed with any physical significance.
On the other hand, the logarithmic UV divergence in \eqref{Kin:corr:wf}
appears to capture some regulator-independent, realistic physical effect,
in which we are most interested.

\section{Relativistic QFT for atoms and OPE: an unsuccessful exploration}
\label{sQED:OPE:failure}

Intuitively, the field-theoretical counterpart of the KG equation in a static potential is just the
complex scalar field theory coupled with an external electrostatic field.
Similar to our preceding work to describe a Schr\"{o}dinger atom,
rather than introduce classical electromagnetic field,
we employ the heavy nucleus effective theory (HNET) which is simply analogous to the
famous heavy quark effective theory~\cite{Eichten:1989zv,Georgi:1990um},
to take into account the quantum fluctuation around a static, positively-charged, and
infinitely-heavy nucleus.
The spin-zero electron coupled with electromagnetism is formulated by the scalar QED (sQED).
The relativistic UV theory that governs all dynamical aspects of atoms is thus formulated
by combining the sQED and HNET,
\beq
{\mathcal L}_{\rm UV} = (D_\mu\phi)^\dagger D^\mu\phi-m^2\phi^\dagger\phi+N^\dagger iD_0 N-\frac{1}{4}F_{\mu\nu}F^{\mu\nu},
\label{UV:atom:lagrangian}
\eeq
where the covariant derivatives acting on electron and nucleus are
\beq
D_{\mu}=\partial_{\mu}+ie A_{\mu},\qquad D_{\mu}=\partial_{\mu}-iZ eA_{\mu},
\eeq
respectively. Equation~\eqref{UV:atom:lagrangian} is obviously gauge invariant. To our purpose, we only need
keep the leading-order term in HNET. Therefore, our formalism has not incorporated the effects of
finite charge radius and magnetic dipole moment of the nucleus,
which are suppressed by powers of $1/M_N$.

Following \cite{Huang:2018yyf}, we may tentatively express the KG wave function of
a hydrogen-like atom in term of the following vacuum-to-atom matrix element~\footnote{In Weinberg's
influential text~\cite{Weinberg:1995mt}, he reinterpreted the Coulomb solution of the Dirac equation
in a field-theoretical context, {e.g.}, in which the corresponding Dirac wave function
is regarded as the vacuum-to-hydrogen matrix element.
He invokes the external field approximation, whose physical foundation very much resembles our HNET approach.
Nevertheless, with a single electron Dirac field operator, it is not obvious to us how to formulate the
OPE in the external field formalism.}:
\beq
\Psi_{nlm}({\rm r}) \equiv \langle 0 \vert  \phi({\bf x}) N(0) \vert n l m \rangle.
\label{KG:wf:nonlocal:matrix:element}
\eeq
Note both field operators are defined in an equal time $t=0$.
Similar to the familiar Bethe-Salpeter wave function of a hadron
in QCD, this spatially-nonlocal matrix element is certainly not gauge invariant.

The universal near-origin behavior of the KG wave function may be
transparently understood from \eqref{KG:wf:nonlocal:matrix:element},
once the Wilson expansion of the products $\phi({\bf x}) N(0)$ is conducted.
In the ${\bf r}\to 0$ limit, the anticipated OPE relation reads
\beq
\phi_R({\bf r}) N_R(0) =  C(r) [\phi N]_R(0)+\cdots,
\label{OPE:sQED:coordinate:space}
\eeq
where the subscript ``$R$'' implies the renormalized operator, and $r\equiv |{\bf r}|$.
The ellipsis indicates all other higher-dimensional operators with less singular Wilson coefficients.
For simplicity, we will only consider the leading $S$-wave operator throughout this work.
It is instructive to also look at the momentum-space version of the OPE relation:
\beq
\widetilde{\phi}_R({\bf q}) N_R(0) \equiv
\int\!\! d^3{\bf{r}}\, e^{-i{\bf{q}}\cdot {\bf{r}}}\phi_R({\bf r}) N_R(0) = \widetilde{C}(q)[\phi N]_R(0)+\cdots,
\label{OPE:sQED:momentum:space}
\eeq
with $q\equiv |{\bf q}|$. This OPE relation is understood to hold true
in the ${\bf q}\to \infty$ limit.

Our goal in the rest of this section is to compute the Wilson coefficients $C(r)$ and $\widetilde{C}(q)$
to the first nontrivial order in $Z\alpha$.
It is curious to examine whether the Wilson coefficient in \eqref{OPE:sQED:coordinate:space}
can reproduce the $\ln r$ term encoded in \eqref{KG:WF:origin:Log:Div},
associated with expansion of the $S$-wave KG wave functions near the origin.

\subsection{Renormalization of local composite operator}
\label{Renorm:Local:Composite:OP:sQED}

Renormalization of local composite operator and OPE are two sides of one coin,
are two concepts closely intertwined with each other.
It is the OPE that endows the renormalized composite operator with
a most unambiguous physical meaning~\cite{Collins:1984xc}.
Technically speaking, all the operators in the right-hand side of the OPE relation
\eqref{OPE:sQED:coordinate:space} must be the renormalized (finite) ones.
So let us first investigate the renormalization of the composite operator $\phi N$.
As usual, the renormalized operator is defined by
\beq
[\phi N]_R(x) \equiv  Z_{e N}\phi^{(0)}(x) N^{(0)}(x) =
Z_{e N}\sqrt{Z_e Z_N} \phi_R(x) N_R(x),
\label{Renormalize:local:composite:op}
\eeq
where the superscript $(0)$ signifies the bare (unrenormalized) quantity.
$Z_e$, $Z_N$, and $Z_{e N}$ are wave function renormalization constants
for the electron field, the nucleus field,
and the composite operator $\phi N$, respectively,
\beq
Z_e=1+\delta_{e},\qquad Z_N=1+\delta_{N},\qquad Z_{e N}=1+\delta_{\phi N}.
\eeq

For later use, we further define a new renormalization constant $Z_S$ through
\beq
[\phi N]_R = Z_S\, \phi_R N_R ,
\label{Renormalize:composite:op:for:OPE}
\eeq
which is connected with $Z_{e N}$ through
\begin{align}
Z_S=1+\delta_{S} =Z_{e N}\sqrt{Z_e Z_N},\qquad \delta_S=\delta_{eN}+{\delta_{e}\over 2}+{\delta_{N}\over 2}.
\end{align}

The renormalization of $\phi N$ is very similar to renormalization of the heavy-light quark
current $\bar{q}\Gamma h_v$ in HQET~\cite{Manohar:2000dt}.
$Z_{e N}$ is electromagnetic gauge invariant only if $Z=1$, for $\phi N$ being electrically neutral.
All the other renormalization constants are gauge-dependent.
In this section, we will choose the Feynman gauge and Coulomb
gauge for quantized electromagnetic field.
The photon propagator in Feynman gauge reads
\beq
D_{\mu\nu}(k)= {-i g_{\mu\nu} \over k^2+i\epsilon},
\eeq
while in Coulomb gauge, its nonvanishing components are
\beq
D_{00}(k)=\frac{i}{{\bf k}^2}, \qquad
D_{ij}(k)= {i\over k^2+i\epsilon} \left(\delta^{ij}-\frac{k^i k^j} {{\bf k}^2}\right),
\quad i,j = 1,2,3.
\eeq
Moreover, as indicated in \eqref{UV:atom:lagrangian}, the
electron and nucleus propagators are simply
\beq
D_e(p) \equiv {i\over p^2-m^2+i\epsilon}, \qquad
D_N(k) \equiv {i \over k^0+i\epsilon}.
\eeq

\begin{figure}[tb]
\centering
\includegraphics[width=0.7\textwidth]{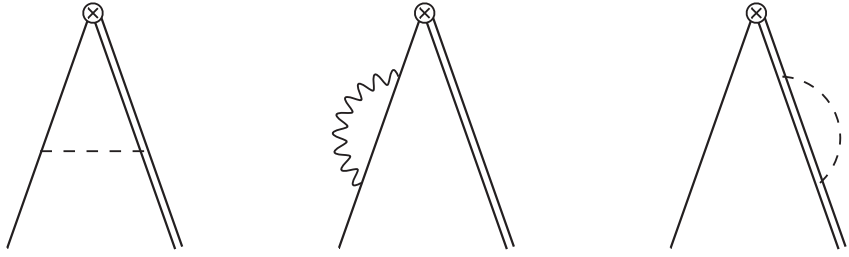}
\caption{The relevant one-loop diagrams for renormalization of the the composite operator $\phi N$
in the relativistic QFT for atoms in \eqref{UV:atom:lagrangian}.
The solid line represents the electron field, the double line represents the HNET field,
whereas the wavy line represents the full photon propagator, and the dashed line
for the temporal component of the photon propagator.
\label{Fig:sQED:local:op:renorm:one:loop}}
\end{figure}

\begin{table}
\begin{tabular}{*5{|c}|}
\hline
&\makecell{$\delta_e$}&\makecell{$\delta_N$}&\makecell{$\delta_S$}&\makecell{$\delta_{e N}$}
\\\hline
\makecell{Feynman}&$\quad\dfrac{\alpha}{\pi\epsilon}\quad$& $\quad\dfrac{Z^2 \alpha}{\pi \epsilon}\quad$&$\quad-\dfrac{Z\alpha}{2\pi\epsilon}\quad$ & $-\dfrac{Z^2\alpha}{2\pi\epsilon}-\dfrac{Z\alpha}{2\pi\epsilon}- \dfrac{\alpha}{2\pi\epsilon}$
\\\hline
\makecell{Coulomb}&$\dfrac{\alpha}{\pi\epsilon}$ &$0$ & $-\dfrac{Z\alpha}{\pi \epsilon}$&$-\dfrac{Z \alpha}{\pi\epsilon}-\dfrac{\alpha}{2\pi\epsilon}$
\\\hline
\end{tabular}
\caption{Numerous counterterms in Feynman gauge and Coulomb gauge in the MS scheme.}
\label{Numerous:Counterterms:sQED}
\end{table}

The corresponding one-loop diagrams for renormalization of $\phi N$
are shown in Fig.~\ref{Fig:sQED:local:op:renorm:one:loop}.
The Feynman gauge results can be adapted from \cite{Manohar:2000dt} with very little modification.
For a gauge theory, it is practically most convenient to utilize the dimensional regularization (DR)
as the UV regulator, which automatically preserves the Poincar\`{e} and gauge invariance.
From now on, we will work in the $D=4-\epsilon$ spacetime dimensions to regularize the UV divergences.
Furthermore, we will employ the minimal subtraction (MS) scheme as the renormalization prescription.

In Table~\ref{Numerous:Counterterms:sQED}, we tabulate various renormalization constants in both gauges~\footnote{
Notice in Coulomb gauge, the nucleus self-energy diagram simply vanishes
because the nucleus can only interact with the $A^0$ field, and
the instantaneous Coulomb propagator does not contain the causal $i\epsilon$ factor,
therefore the contour integral yields a vanishing result since
the HNET propagator only contains a single pole in the complex $k^0$ plane.}.
One can readily verify $Z_{eN}= 1- {3 \alpha\over 2\pi\epsilon}$ is indeed gauge-independent
if $Z=1$.

To evaluate $Z_S$ defined in \eqref{Renormalize:composite:op:for:OPE}, one
only needs consider the vertex correction diagram
in Fig.~\ref{Fig:sQED:local:op:renorm:one:loop}, which
turns out to be logarithmically UV divergent.
The counterterm $\delta_S$ thus does not vanish at one-loop order.
As recorded in Table~\ref{Numerous:Counterterms:sQED},
the specific value of $\delta S$ is gauge-dependent, which
equals
\beq
Z_S\Big{\vert}_{\rm Feyn} = 1-{Z\alpha\over 2\pi}{1\over \epsilon}, \qquad
Z_S\Big{\vert}_{\rm Coul} = 1-{Z\alpha\over \pi}{1\over \epsilon},
\label{ZS:in:both:gauges}
\eeq
in Feynman and Coulomb gauge, respectively.
Therefore,  the Wilson coefficient $C({\bf r})$
in \eqref{OPE:sQED:coordinate:space} is anticipated to begin with $Z\alpha \ln r$, in conflict with
the factor $Z^2\alpha^2 \ln r$ in \eqref{KG:Universal:Small:r} for the $S$-wave KG wave function.
This is a warning sign that we are on the wrong track.

\subsection{Wilson expansion of the product of $\bm\phi$ and $\bm{N}$}

The remarkable merit of operator product expansion is that it holds at the operator level.
It does not really matter which specific form of Green function or matrix element should be taken
in order to formulate the Wilson expansion.
For simplicity, we will consider the following four-point Green functions:
\bseq
\bqa
\Gamma({\bf{r}};p,k) & \equiv &
\int \! d^4y \, d^4z\  e^{-ip\cdot y - ik\cdot z}
\langle 0 |T \left\{ \phi({\bf r}) N(0)\phi^\dagger (y) N^\dagger (z)\right\}|0\rangle_{\rm amp},
\label{Four-point:Green:function:coordinate}
\\
\widetilde{\Gamma}({\bf{q}};p,k) & \equiv &
\int \! d^4y \, d^4z\  e^{-ip\cdot y-ik\cdot z} \langle 0 |T \left\{ \widetilde{\phi}({\bf q}) N(0)\phi^\dagger (y) N^\dagger (z)\right\}|0\rangle_{\rm amp},
\label{Four-point:Green:function:momentum}
\eqa
\label{Def:two:amputated:Green:functions}
\eseq
in the coordinate space and momentum space, respectively.
Here $p$, $k$ can be arbitrary (off-shell) ``soft'' momenta~\footnote{Some quantification on the term ``soft'' in
the context of HNET may be necessary. In HQET, heavy quarks can only interact with
the soft degrees of freedom, those with momentum $k^\mu \sim \Lambda_{\rm QCD}\ll m_Q$,
where $\Lambda_{\rm QCD}$ signifies the intrinsic nonperturbative hadronic scale, and $m_Q$
implies the heavy quark mass.
There is no counterpart of $\Lambda_{\rm QCD}$ in QED. Thus one may regard any momentum scale, {\it e.g.},
the electron mass, to be ``soft'', provided that it is much smaller than the nucleus mass $M_N$.
In this work, we are considering the infinite-nucleus-mass limit, thereby any finite momentum scale
can be treated as ``soft'' in this sense.}.
For definiteness, one may simply assume $p, k\sim m$.
The subscript ``amp'' in \eqref{Def:two:amputated:Green:functions} implies that the
external propagators carrying soft momenta $p$ and $k$ get amputated.
Two Green functions in \eqref{Def:two:amputated:Green:functions} are related to each other
through Fourier transform:
\beq
\widetilde{\Gamma}({\bf{q}}; p, k ) =
\int\!\! d^3{\bf{r}}\, e^{-i{\bf{q}}\cdot {\bf{r}}} \,\Gamma({\bf{r}};\,p, k ),\qquad
 \Gamma({\bf{r}};\,p, k )= \int\!\! {d^3{\bf{q}} \over (2\pi)^3}\, e^{i{\bf{q}}\cdot {\bf{r}}} \,\widetilde{\Gamma}({\bf{q}};p, k ).
\label{Gamma:Fourier:transform}
\eeq

For convenience, we also define the three-point Green function
with the renormalized local operator $[\phi N]_R$ inserted:
\beq
G(p, k)  \equiv
\int \! d^4y \, d^4z\  e^{-ip\cdot y-ik\cdot z}
\langle 0 |T \left\{ [\phi N]_R(0) \phi^\dagger (y) N^\dagger(z) \right\}|0\rangle_{\rm amp},
\label{Three-point:Green:function}
\eeq

The OPE can be verified by ensuring the four-point Green functions do bear the following factorized form,
order by order in $Z\alpha$:
\beq
\Gamma({\bf{r}};p,k)  \xrightarrow{q\gg m} C(r) G(p, k)+\cdots, \qquad
\widetilde{\Gamma}({\bf{q}};p,k) \xrightarrow{q\gg m} \widetilde{C}(q) G(p, k)+\cdots.
\label{OPE:sQED:verification}
\eeq

\begin{figure}[tb]
\centering
\includegraphics[width=0.5\textwidth]{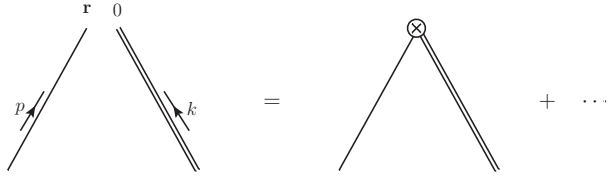}
\caption{The leading term in the expansion for the product of $\phi$ and $N$ fields in the free theory.
\label{Fig:free:theory:ope}}
\end{figure}

As indicated in Fig.~\ref{Fig:free:theory:ope}, the lowest-order prediction of the four-point Green function
in the coordinate space comes from the free theory:
\beq
\Gamma^{(0)}({{\bf r}};p,k)= e^{i {\bf p} \cdot{\bf r}} = 1 + {\cal O}(p r),
\label{LO:Prediction}
\eeq
just arising from the naive Taylor expansion. Since $G^{(0)}(p, k)=1$, we can read off the leading
Wilson coefficient to be $C(r)=1+{\cal O}(Z\alpha)$.
It corresponds to $\widetilde{\Gamma}^{(0)}({{\bf q}};p,k)=(2\pi)^3 \delta^3({\bf q})$, therefore
$\widetilde{C}(q)= (2\pi)^3\delta^3({\bf q})+{\cal O}(Z\alpha)$.

\begin{figure}[tb]
\centering
\includegraphics[width=1.0\textwidth]{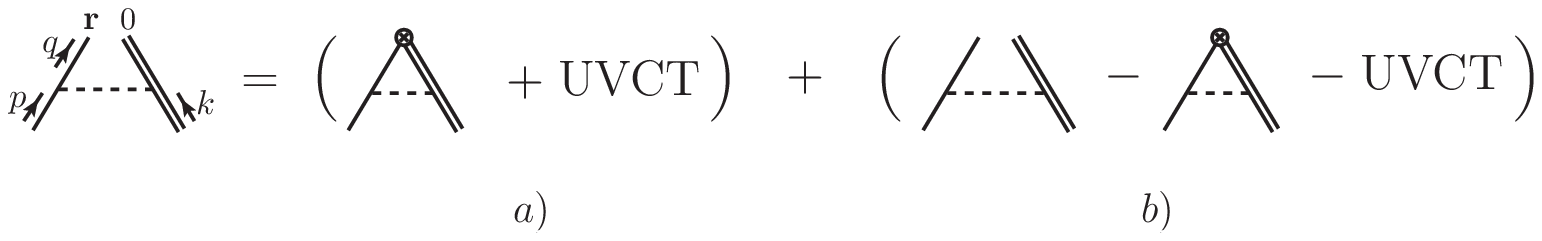}
\caption{The next-to-leading term in the expansion for the product of $\phi$ and $N$ fields in the coordinate space.
The original four-point Green function is reexpressed as the sum of two terms,
by subtracting and adding back the renormalized Green function inserted with a local $\phi N$ operator.
\label{Fig:NLO:OPE:coordinate:space}}
\end{figure}

Next we consider the correction to the Wilson coefficients after inclusion of the interaction.
We may express these coefficients as
\beq
C(r)=1+c_1(r)+\cdots, \qquad \widetilde{C}(q)= (2\pi)^3\delta^{(3)}({\bf q})+ \tilde{c}_1(q)+\cdots,
\label{Wilson:coefficients:Exp:SQED:HNET}
\eeq
where $c_1(r)$ and $\tilde{c}_1(q)$ are the Wilson coefficients at the order $Z\alpha$.

Let us first analyze the large-$q$ behavior of the four-point Green function in momentum space
in \eqref{Four-point:Green:function:momentum}, as depicted by the leftmost
diagram in Fig.~\ref{Fig:NLO:OPE:coordinate:space}.
A short calculation in Feynman gauge leads to
\bqa
&& \widetilde{\Gamma}^{(1)}({\bf{q}}; p,k)\Big|_{\rm Feyn}  =
Ze^2 \int \!\! {dq^0\over 2\pi}\, (q^0+p^0) D_e(q) D_{00}(q-p) D_N(k+p-q)
\nn\\
&& =  -\frac{Ze^2}{2} \Bigg[\frac{\sqrt{{\bf q}^2+m^2}+p^0}{\sqrt{{\bf q}^2+m^2}(\sqrt{{\bf q}^2+m^2}-k^0-p^0)\left((\sqrt{{\bf q}^2+m^2}-p^0)^2-({\bf q-p})^2\right)}
\nn\\
&&  + \frac{2p^0+|{\bf q-p}|}{(|{\bf q-p}|-k^0)|{\bf q-p}|((p^0+|{\bf q-p}|)^2-{\bf q}^2-m^2)}\Bigg],
\label{momentum:4-point:Green:Feyn:gauge}
\eqa
where the standard contour method is utilized.
Expanding \eqref{momentum:4-point:Green:Feyn:gauge} around the
$q\to \infty$ separately, one finds that the leading $1/|{\bf q}|$ term cancels upon summing the two terms inside
the bracket, and the net limiting behavior turns out to scale as only $1/|{\bf q}|^3$:
\beq
\widetilde{\Gamma}^{(1)}({\bf{q}}; p,k)\Big|_{\rm Feyn} \xrightarrow{q\gg m} \frac{\pi Z\alpha}{|{\bf q}|^3}+\cdots.
\label{Asym:behav:momentum:4-point:Green:Feyn:gauge}
\eeq
Using $G^{(0)}(p, k)=1$, we thus verify that $\widetilde{\Gamma}$ does possess
the desired OPE texture, and obtain
\beq
\tilde{c}_1(q)\big|_{\rm Feyn} =\frac{\pi Z\alpha}{|{\bf q}|^3}.
\label{WC:sQED:c1q:Feyn}
\eeq

The same analysis can be repeated in Coulomb gauge, resulting in simpler form.
The corresponding momentum-space Green function is
\bqa
\widetilde{\Gamma}^{(1)}({\bf{q}}; p,k)\Big|_{\rm Coul}  &=&
-\frac{Ze^2}{2}\frac{p^0+\sqrt{{\bf q}^2+m^2}}{\sqrt{{\bf q}^2+m^2}(k^0+p^0-\sqrt{{\bf q}^2+m^2})({\bf q-p})^2}
\nn\\
& \xrightarrow{q\gg m}&\frac{2\pi Z\alpha}{|{\bf q}|^3}+\cdots.
\label{momentum:4-point:Green:Coulomb:gauge}
\eqa
Thus we have
\beq
\tilde{c}_1(q)\big|_{\rm Coul} = {2\pi Z\alpha\over |{\bf q}|^3}.
\label{WC:sQED:c1q:Coul}
\eeq

When Fourier-transforming $\widetilde{\Gamma}^{(1)}$ into $\Gamma^{(1)}$ in accord with
\eqref{Gamma:Fourier:transform}, the integration variable ${\bf q}$ can range from soft to hard.
If $|{\bf q}|\sim p, k, m$, the characteristic soft regime,
it is legitimate to approximate the exponential $e^{i {\bf q}\cdot {\bf r}}$ by 1.
Supplemented with the UV counterterm, this piece of contribution is captured by the renormalized Green function,
as represented by Fig.~\ref{Fig:NLO:OPE:coordinate:space}$a)$.
On the contrary, when $q$ is in the hard regime, {\it i.e.}, $|{\bf q}|\gg m\sim 1/r$,
one can no longer expand the Fourier exponential $e^{i {\bf q}\cdot {\bf r}}$.
But the integration over the hard region should be encapsulated in the Wilson coefficient.
Precisely speaking, the Wilson coefficient can then be identified with the original coordinate-space Green
function $\Gamma^{(1)}({\bf r})$ subtracting off the Green function containing the renormalized $\phi N$ operator:
\beq
c_1(r) = \lim_{r\to 0} \Gamma^{(1)}({\bf{r}}; p,k)- G_R^{(1)}(p,k).
\label{OPE:sQED:c1:extract}
\eeq
This add-and-subtract procedure has been pictured in Fig.~\ref{Fig:NLO:OPE:coordinate:space},
which can also be derived from \eqref{OPE:sQED:verification}.

As represented in Fig.~\ref{Fig:NLO:OPE:coordinate:space}$a)$,
the NLO renormalized Green function with the $\phi N$ insertion,
$G_R^{(1)}$, is defined by
\beq
G_R^{(1)}(p,k;\mu)= \mu^\epsilon \int\!\! {d^{D-1} {\bf{q}} \over (2\pi)^{D-1}} \widetilde{\Gamma}^{(1)}({\bf{q}};p, k )+\delta_S,
\label{Renormalized:Green:function:sQED}
\eeq
where the UV counterterm $\delta_S$ for both gauges are given in
Table~\ref{Fig:sQED:local:op:renorm:one:loop} as well as \eqref{ZS:in:both:gauges}.
The MS scheme has been utilized to renormalize the
Green function $G^{(1)}$.

In accordance with Fig.~\ref{Fig:NLO:OPE:coordinate:space}$b)$,
making use of \eqref{Gamma:Fourier:transform} and \eqref{Renormalized:Green:function:sQED},
we reexpress $c_1(r)$ in \eqref{OPE:sQED:c1:extract} as
\bqa
c_1(r) &=& \mu^{\epsilon}\int\!\!
{d^{D-1}{\bf q}\over (2\pi)^{D-1}}  \widetilde{\Gamma}^{(1)}({\bf{q}};p, k )
\left( e^{i{\bf{q}}\cdot {\bf r}}-1\right)-\delta_S
\nn \\
& \approx &\mu^{\epsilon}\int {d^{D-1}{\bf q}\over (2\pi)^{D-1}}\,
\tilde{c}_1(q) \left(e^{i{\bf{q}}\cdot {\bf r}}-1\right)-\delta_S.
\label{c1r:Fourier:trans:from:c1q}
\eqa
In the second line, we have only kept the leading asymptotic form of $\widetilde{\Gamma}^{(1)}$ in the
$q\to \infty$ limit. Note the subtraction form automatically guarantees that, upon dropping all
the occurrence of the $p$, $k$ and $m$, the infrared divergence that may potentially arise in the
${\bf q}\to 0$ limit does not actually arise! By construction, $c_1(r)$ in \eqref{c1r:Fourier:trans:from:c1q} is
both UV and IR finite, as it must be.

Substituting \eqref{WC:sQED:c1q:Feyn} and \eqref{WC:sQED:c1q:Coul} into \eqref{c1r:Fourier:trans:from:c1q},
it is straightforward to obtain the intended order-$Z\alpha$
coordinate-space Wilson coefficients in both gauges:
\beq
c_1(r)\big|_{\rm Feyn} =  -\dfrac{Z\alpha}{2\pi}\left(\ln{\mu r}+{\gamma_E\over 2} + \ln \pi \right),
\qquad
c_1(r)\big|_{\rm Coul} = -\dfrac{Z\alpha}{\pi}\left(\ln{\mu r}+{\gamma_E\over 2} + \ln \pi \right),
\eeq
where $\gamma_E=0.5772\cdots$ is the Euler gamma constant.
As expected, the coefficients of the $\ln r$ term in two gauges are identical to the
coefficient of the single pole in $\delta_S$ in \eqref{ZS:in:both:gauges}.

Albeit being gauge-dependent, the Wilson coefficient $C(r)$ in \eqref{OPE:sQED:coordinate:space}
indeed contains the $\ln r$ term, which however arises already at order $Z\alpha$.
This is in sharp contrast with the weak logarithmic singularity of the $S$-wave KG wave functions
near the origin encapsulated in \eqref{KG:WF:origin:Log:Div},
in which the $\ln r$ term is accompanied with a coefficient $\propto Z^2\alpha^2$.

Therefore, we must admit that the Wilson expansion based on relativistic scalar QED fails to
account for the divergence symptom of KG wave function near the origin.
It may casts some doubt on the legitimateness of interpreting the matrix element
\eqref{KG:wf:nonlocal:matrix:element} as the KG wave function,
which somewhat contradicts the underlying assumption
in Weinberg's text on Dirac equation in Coulomb potential~\cite{Weinberg:1995mt}.
In particular, the short-distance ($r\ll 1/m$) behavior of KG wave function cannot be correctly accounted by
the OPE formalism described in this section. As a superficially relativistic wave equation,
incorporation of the instantaneous Coulomb potential forces us to choose a special reference frame for the
KG equation, namely the rest frame of the nucleus.
Since the simultaneous production  of the electron-positron pair becomes inevitable once the distance is probed in
a distance shorter than Compton wavelength, the single-particle probabilistic interpretation of relativistic
wave equation must breakdown. The lesson seems to be that one should not attempt to discuss the
very short-distance behavior of the KG equation, which might not bear enough physical significance.

\section{A Proper nonrelativistic effective field theory for atom}
\label{Proper:NREFT}

In the preceding section, having started with a relativistic QFT to describe the atom,
which is based on the sQED plus HNET,
we have gone through a long way to prove the OPE along this direction actually leads to a dead end.
Nevertheless, we recall that in section~\ref{Quantum:perturb:corr:wave:func:origin},
upon a nonrelativistic reduction of KG equation into an effective Schr\"{o}dinger equation,
one is able to embed the relativistic effects into a few corrections to the Coulomb Hamiltonian.
The kinetic correction term appears to capture the correct structure of the
short-range logarithm by employing the quantum-mechanical perturbation
theory to analyze the first-order correction to the Schr\"{o}dinger wave function at the origin.
This is a very encouraging result, which insinuates the correct way is to
formalize the OPE in the framework of a non-relativistic
effective field theory framework, rather than the relativistic QFT considered in section~\ref{sQED:OPE:failure}.

The original KG equation in \eqref{KG:Equation} only involves the Coulomb interaction between the electron and nucleus, since that
equation is specified in the rest frame of the nucleus.
To caputure this feature, we drop all the occurrences of the vector potential ${\bf{A}}$ in \eqref{UV:atom:lagrangian}.
The simplified UV field theory for atom now reads
\beq
{\cal L}_{\rm UV^\prime} = (D_0\phi)^\dagger D^0\phi-{\bf{\nabla}}\phi^\dagger\cdot
{\bf \nabla}\phi-m^2\phi^\dagger\phi+N^\dagger iD_0N+\frac{1}{2}(\bm{\nabla}A_0)^2.
\label{Lag:full:sQED:without:vec:A}
\eeq

Eq.~\eqref{Lag:full:sQED:without:vec:A} is certainly no longer gauge invariant.
However, this does not necessarily imply a nuisance. Coulomb gauge $\nabla \cdot {\bf A}=0$ is a particularly
convenient choice for tackling nonrelativistic bound state problem,
where the effect of $\bf A$ and $A^0$ can be cleanly separated. The former encodes the dynamics of the transversely polarized
photon, whose effect is absent in this work; the latter is the very agent for mediating the instantaneous Coulomb interaction,
which provides the crucial binding mechanism for all atoms. From now on, we will stay exclusively with the Coulomb gauge
in the rest of the work~\footnote{A common representation of \eqref{Lag:full:sQED:without:vec:A} it to
get rid of $A^0$ and in favor of explicit introducing the instantaneous Coulomb potential.
Nevertheless, we stress that \eqref{Lag:full:sQED:without:vec:A} is still a {\it local} field theory. Note
locality plays an essential role in corroborating OPE,
in accord with Weinberg's proof of OPE using path integral~\cite{Weinberg:1996kr}.}.

\subsection{NREFT from field redefinition}
\label{field:redefinition}

Analogous to deriving the effective Hamiltonian in \eqref{Effective:Schr:Hamiltonian}
from the single-particle relativistic KG equation,
our goal is to present a quick construction of the nonrelativistic EFT,
starting from the scalar QED sector of \eqref{Lag:full:sQED:without:vec:A},
\beq
{\mathcal L}_{\rm sQED^\prime}= (D_0\phi)^\dagger D^0\phi-c^2(\bm{\nabla}\phi)^\dagger\cdot\bm{\nabla}\phi-m^2c^4\phi^\dagger\phi,
\label{Lag:sQED:without:vec:A}
\eeq
with the corresponding Euler-Lagrange equation reading
\beq
\left( D_0^2- c^2 {\bm\nabla}^2+ m^2 c^4\right)\phi=0.
\label{sQED:equation:of:motion}
\eeq

To separate the electron and positron degrees of freedom, we first rewrite $\phi$ as~\footnote{The
field redefinition \eqref{phi:seperation:two:pieces} is employed to derive the heavy scalar effective theory (HSET)
lagrangian to sub-leading order in $1/m$ expansion~\cite{Schwartz:2013pla}.
Here we use this technique to derive the nonrelativistic sQED (NRsQED) lagrangian
to order $v^4/c^4$.}
\beq
\phi={1\over \sqrt{2m c^2}} e^{-im c^2 t}(\varphi+\tilde{\varphi}),
\label{phi:seperation:two:pieces}
\eeq
with
\beq
\varphi=e^{im  c^2 t}{1\over \sqrt{2m c^2}}(iD_0+m c^2)\phi,\qquad
\tilde{\varphi}=e^{im c^2 t}{1\over \sqrt{2m c^2}}(-iD_0+m c^2)\phi,
\label{Def:variphi:tilde:varphi}
\eeq
where $\varphi$ ($\tilde{\varphi}$) may be referred to as the large (small) component.
Substituting \eqref{phi:seperation:two:pieces} into \eqref{Lag:sQED:without:vec:A}, we then obtain
\beq
{\mathcal L}_{\rm sQED^\prime}=\left(\varphi^\dagger +\tilde{\varphi}^\dagger\right)
\left(iD_0+\frac{\bm{\nabla}^2}{2m}-\frac{D_0^2}{2mc^2}\right) (\varphi+\tilde{\varphi}),
\label{lag:sQED:alternative}
\eeq
where integration by part is used to make all differential operators act to the right.

Acting the covariant derivative $-iD^0$ on $\varphi+\tilde{\varphi}$, with the aid of
\eqref{phi:seperation:two:pieces} and \eqref{Def:variphi:tilde:varphi},
we can find the relation
\begin{align}
\tilde{\varphi}=-{1\over 2m c^2} i D_0\left(\varphi+\tilde{\varphi}\right).
\label{D0:varphi:plus:tildevarphi}
\end{align}
Assuming $iD_0\sim mv^2$, one infers from \eqref{D0:varphi:plus:tildevarphi} that
$\tilde{\varphi}\approx -{1\over 2m c^2} i D_0 \varphi \sim {\cal O} (v^2/c^2) \varphi$,
thus justifying the nomenclature ``small'' component for $\tilde{\varphi}$.

Substituting \eqref{phi:seperation:two:pieces} into \eqref{sQED:equation:of:motion}, or directly
starting from the lagrangian \eqref{lag:sQED:alternative}, we obtain the
following equation of motion for $\varphi$ and $\tilde{\varphi}$:
\beq
\left[iD_0 + {\bm{\nabla}^2\over 2m} - {D_0^2\over 2mc^2} \right](\varphi+\tilde{\varphi})=0.
\label{varphi:tildevarphi:EOM}
\eeq

Utilizing \eqref{D0:varphi:plus:tildevarphi} to get rid of $i D_0(\varphi+\tilde{\varphi})$ in
\eqref{varphi:tildevarphi:EOM}, after some straightforward algebra, one finds
\begin{align}
iD_0 \varphi= {1 \over 2m} {\bm\nabla}^2\left(\varphi+\tilde{\varphi}\right).
\label{D0:varphi:identity}
\end{align}

Now we are ready to derive the nonrelativistic scalar QED lagrangian. First we retain the $\varphi^\dagger(iD_0+\frac{\bm{\nabla}^2}{2m})\varphi$ term in \eqref{lag:sQED:alternative}
as the minimal NRsQED lagrangian. We the apply equations \eqref{D0:varphi:plus:tildevarphi} and \eqref{D0:varphi:identity}
iteratively, to get rid of the occurrence of $\tilde{\varphi}$ consecutively order by order in $v$ expansion.
To be explicit, let us take the ${\cal O}(v^4/c^4)$ term $\tilde{\varphi}^\dagger(iD_0)\tilde{\varphi}$ as an explicit example
to illustrate our procedure:
\bqa
\tilde{\varphi}^\dagger iD_0 \tilde{\varphi} & \approx &
\left(-\frac{1}{2mc^2}iD_0 \varphi\right)^\dagger iD_0 \left(-\frac{1}{2mc^2} iD_0\varphi\right) \approx
{1\over 16 m^4 c^4}(\bm{\nabla}^2 \varphi)^\dagger iD_0 \bm{\nabla}^2 \varphi
\nn\\
&=& {1\over 16 m^4 c^4} (\bm{\nabla}^2 \varphi)^\dagger \left(\bm{\nabla}^2 iD_0+
[iD_0,\bm{\nabla}^2]\right)\varphi
\nn\\
&\approx &
-{1\over 32m^5c^4}\varphi^\dagger {\bm \nabla}^6\varphi
  -{1\over 16m^4c^4} \varphi^\dagger {\bm \nabla}^2[\bm{\nabla}^2,e A_0]\varphi.
\label{subexp}
\eqa
In deriving this, we have used \eqref{D0:varphi:plus:tildevarphi},
\eqref{D0:varphi:identity} and the integration by part in the last step.

After some straightforward yet tedious iterative derivations,
we arrive at the NRsQED lagrangian accurate to relative order $v^4/c^4$,
\beq
\mathcal{L}_{\rm NRsQED}=\varphi^\dagger(iD_0+\frac{\bm{\nabla}^2}{2m}+\frac{\bm{\nabla}^4}{8m^3c^2}+
\frac{\bm{\nabla}^6}{16m^5c^4}+\frac{1}{16m^4c^4}\bm{\nabla}^2[\bm{\nabla}^2,eA^0])\varphi.
\label{eff:lagrangian:sNRQED}
\eeq
The last term is not in the right form of the Darwin term as given
in \eqref{DeltaH:perturb}. One may supplement the effective lagrangian \eqref{eff:lagrangian:sNRQED}
with the following total time derivative term:
\bqa
{1\over 32 m^4 c^4} i\partial_0 (\varphi^\dagger\bm{\nabla}^4 \varphi)
&=& -{1\over 64m^5c^4} \bm{\nabla} \cdot \left[(\bm{\nabla}\varphi)^\dagger \bm{\nabla}^4\varphi-\varphi^\dagger \bm{\nabla}^4(\bm{\nabla}\varphi)\right]
\\
&+& {1\over 32m^4c^4}
\varphi^\dagger\left[\bm{\nabla}^4e A_0- eA_0\bm{\nabla}^4\right]\varphi,
\nn
\eqa
where \eqref{D0:varphi:identity} is used.

Now the nonrelativistic sQED is put in the canonic form.
Together with the HNET and the reduced Maxwell lagrangian,
we finally write down our nonrelativistic EFT for atoms:
\bqa
{\cal L}_{\rm NREFT}&= & \varphi^\dagger
\left(iD_0+\frac{\bm{\nabla}^2}{2m}\right) \varphi + \varphi^\dagger \frac{\bm{\nabla}^4}{8m^3} \varphi+
\varphi^\dagger \frac{\bm{\nabla}^6}{16m^5} \varphi + {1\over 32m^4} \varphi^\dagger [\bm{\nabla}^2,[eA_0,\bm{\nabla}^2]]\varphi
\nn\\
&+& N^\dagger i D_N^0 N +\frac{1}{2}(\bm{\nabla}A_0)^2 + {c_4\over m^2} \varphi^\dagger\varphi N^\dagger N +  \cdots,
\label{NREFT:spin-0-electron:atom}
\eqa
where we have recovered natural unit. In the NRsQED sector, we retain all the allowed terms
through the relative order $v^4$.
Apart from rotational invariance, this NREFT also obeys a global
$U(1)\times U(1)$ phase symmetry, which reflects the separate number conservation law for electrons and nucleus,
respectively. Rotation and phase invariance also allows us to add a contact interaction $\varphi^\dagger\varphi N^\dagger N$
to \eqref{NREFT:spin-0-electron:atom}.
The corresponding dimensionless coefficient $c_4$ encodes some possible
short-distance dynamics.

We emphasize that, as a nonrelativistic effective theory, \eqref{NREFT:spin-0-electron:atom} is only
valid when the probed momentum is smaller than its intrinsic UV cutoff, which is of order electron mass $m$.

\subsection{Tree-Level Matching}
\label{tree:level:matching}

\begin{figure}[tb]
\centering
\includegraphics[width=0.7\textwidth]{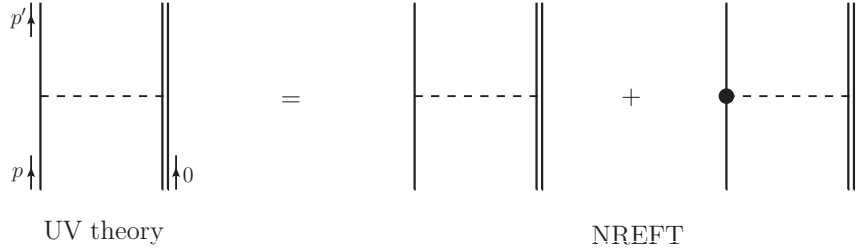}
\caption{The tree-level matching from the UV theory onto the NREFT.
The heavy dot represents the Darwin term.
\label{Fig:sQED:tree:matching}}
\end{figure}

Let us use the more standard perturbative matching method to
verify our NREFT lagrangian in \eqref{NREFT:spin-0-electron:atom}.
The strategy is familiar, to demand both UV theory  \eqref{Lag:full:sQED:without:vec:A} and
EFT \eqref{NREFT:spin-0-electron:atom} yield equal $S$-matrix for the elastic scattering
$e(p)N\to e(p^\prime)N$ in the low-energy limit, order by order
in loop and velocity expansion.

We first consider matching the tree-level amplitude, which is illustrated in Fig.~\ref{Fig:sQED:tree:matching}.
The UV theory generates the amplitude:
\beq
i \mathcal{M}^{\rm tree}_{\rm UV}= \frac{i Ze^2 }{\sqrt{2p^0}\sqrt{2{p'}^0}}\, {p^0+{p'}^0\over ({\bf p}-{\bf p'})^2}
=\frac{i Z e^2}{({\bf p}-{\bf p'})^2}+
\frac{iZe^2({\bf p}^4-2{\bf p}^2{\bf p'}^2+{\bf p'}^4)}{32m^4({\bf p}-{\bf p'})^2}+\mathcal{O}(v^5).
\label{tree:ampl:sQED:HNET}
\eeq
We have adopted the nonrelativistic normalization convention for states,
which is reflected in the prefactor ${1\over \sqrt{2p^0}\sqrt{2{p'}^0}}$.
In deriving the final expression in \eqref{tree:ampl:sQED:HNET},
we have expanded the energies of the incoming and outgoing electron according to
\beq
p^0 = \sqrt{{\bf p}^2+m^2}=m\left(1+\frac{{\bf p}^2}{2m^2}-\frac{{\bf p}^4}{8m^4}
+\cdots\right).
\eeq

As instructed by Fig.~\ref{Fig:sQED:tree:matching}, if one includes the Coulomb interaction
together with the Darwin term in the EFT side, it is then straightforward to verify that
the resulting tree-level NREFT amplitude exactly coincides with \eqref{tree:ampl:sQED:HNET}.

\subsection{One-loop Matching and absence of contact interaction}
\label{one-loop:level:matching}

\begin{figure}[tb]
\centering
\includegraphics[width=1.0\textwidth]{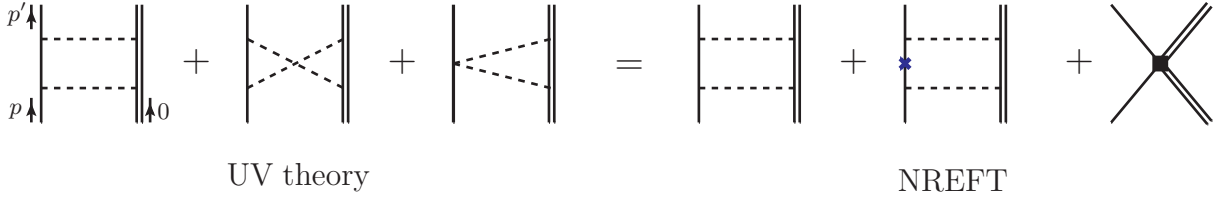}
\caption{One-loop diagrams for the $eN\to eN$ amplitude in both UV theory and NREFT.
The cross represents the ${\bf p}^4$ kinetic correction, and the solid square represents the $c_4$ electron-nucleus
contact interaction.
\label{Fig:sQED:one-loop:matching}}
\end{figure}

We continue to match the low-energy scattering amplitude for $eN\to eN$
from the UV theory onto NREFT at order $Z^2\alpha^2 v^0$,
with the corresponding diagrams shown in Fig.~\ref{Fig:sQED:one-loop:matching}.
The goal is to determine the Wilson coefficient $c_4$ for the contact interaction,
which may encapsulate some effects related to the short-lived
electron-positron intermediate state.

On the UV theory side, the seagull diagram does not contribute, since
only a single pole is present in the integrand that originates from the HNET propagator,
thereby simply generates vanishing result upon contour integral.
Therefore we only need consider the box and crossed box diagrams,
\bqa
\label{One:loop:UV:theory:amplitude}
& &  i \mathcal{M}^{\rm one-loop}_{\rm UV}
 = {Z^2e^4 \mu^{2\epsilon} \over \sqrt{2p^0}\sqrt{2{p'}^0}}\int\!\! \frac{d^D l}{(2\pi)^D} D_e(l)D_{00}(p-l)D_{00}(l-p')
\nn\\
&& \qquad\qquad\quad \times \left[D_N(p-l)+D_N(l-p')\right](p^0+l^0)(l^0+{p'}^0)
\\
&&= {iZ^2e^4  \mu^{2\epsilon} \over \sqrt{2p^0}\sqrt{2p'^0}}\int\!\!
{d^{D-1}{\bf l}\over (2\pi)^{D-1}}{1\over({\bf p}-{\bf l})^2({\bf p'}-{\bf l})^2 2\sqrt{{\bf l}^2+m^2}}
\left(\frac{(p^0+\sqrt{{\bf l}^2+m^2})^2}{\sqrt{{\bf l}^2+m^2}-p^0}+\frac{(p^0-\sqrt{{\bf l}^2+m^2})^2}{\sqrt{{\bf l}^2+m^2}+p'^0}\right)
\nn\\
&&=  {iZ^2e^4  \mu^{2\epsilon} \over \sqrt{2p^0}\sqrt{2p'^0}}\int \!\! {d^{D-1}{\bf l}\over (2\pi)^{D-1}}
{ 4m^2+ 3{\bf p}^2+{\bf l}^2 \over ({\bf l}-{\bf p})^2({\bf l}-{\bf p}^\prime)^2 {  ({\bf l}^2-{\bf p}^2) }  },
\nn
\eqa
where the on-shell condition $p^0=p'^{0}=\sqrt{{\bf p}^2+m^2}$ has been used.

Next we turn to the calculation on the EFT side.
Note the electron propagator in NREFT obeys the nonrelativistic dispersion relation,
\beq
{\cal D}_e(k)\equiv\frac{i}{k^0-\frac{{\bf k}^2}{2m}+i\epsilon}.
\eeq
Accordingly, the order-$Z^2\alpha^2$ NREFT amplitude for $eN\to eN$ becomes
\bqa
\label{One:loop:NREFT:amplitude}
&& i\mathcal{M}^{\rm one-loop}_{\rm NREFT}
=Z^2 e^4\mu^{2\epsilon}\int\!\! \frac{d^D l}{(2\pi)^D} {\cal D}_e(l) D_{00}(p-l)D_{00}(l-p')
D_N(p-l)\left[1+{\cal D}_e(l)\frac{i{\bf l}^4}{8m^3}\right]+ {c_4\over m^2},
\nn\\
&=&iZ^2e^4 \mu^{2\epsilon}\int \!\! \frac{d^{D-1}{\bf l}}{(2\pi)^{D-1}}\frac{1}{({\bf l}-{\bf p})^2({\bf l}-{\bf p}')^2
\left(\frac{{\bf p}^2}{2m}-\frac{{\bf l}^2}{2m}\right)}\left(-1+\frac{{\bf l}^4}{8m^3\left(\frac{{\bf p}^2}{2m}-\frac{{\bf l}^2}{2m}\right)}\right)
+ {c_4\over m^2}.
\eqa
where we have also included the diagram with one ${\bf p}^4$
kinetic term inserted  on the electron propagator.

In passing, we emphasize some remarkable calculational simplifications in NREFT
compared with the relativistic UV theory.
Due to the peculiar causal structure of the nonrelativistic propagators,
and the non-propagating nature of the instantaneous Coulomb interaction,
all diagrams of the crossed-ladder topology, or of the structure with the emission and absorbtion of a Coulomb photon
by the same nonrelativistic line, will all yield vanishing results upon the contour integration over the
$0$-th component of the loop momentum. Therefore, in the rest of the paper,
we need retain those diagrams only exchanging the instantaneous Coulomb ladders.

By construction, \eqref{One:loop:UV:theory:amplitude} and \eqref{One:loop:NREFT:amplitude} must
share the same IR/nonanalytic behavior. They can only differ in the ultraviolet regime
of the loop integral as ${\bf l}\sim m$. This difference may be compensated by
inclusion of the contact interaction in the NREFT. For the sake of deducing $c_4$, it is legitimate
to expand the integrands in \eqref{One:loop:UV:theory:amplitude} and \eqref{One:loop:NREFT:amplitude}
in power series of ${\bf p}/{\bf l},\,{\bf p}^\prime/{\bf l}$,
since the Wilson coefficient receives the contribution entirely from the hard loop momentum regime.

Since the contact term corresponds to $S$-wave operator,
it is feasible to set ${\bf p}={\bf p}^\prime=0$ in \eqref{One:loop:UV:theory:amplitude} and
\eqref{One:loop:NREFT:amplitude}. We then obtain
\bseq
\bqa
&& i \mathcal{M}^{\rm one-loop}_{\rm Full}  =
i Z^2e^4 \mu^{2\epsilon} \int \!\! {d^{D-1}{\bf l}\over (2\pi)^{D-1}}
\left({2m\over {\bf l}^6}+ {1\over 2 m {\bf l}^4}\right)+ \cdots,
\\
&& i\mathcal{M}^{\rm one-loop}_{\rm NREFT} = i Z^2e^4  \mu^{2\epsilon} \int \!\! {d^{D-1}{\bf l}\over (2\pi)^{D-1}}
\left({2m\over {\bf l}^6}+ {1\over 2 m {\bf l}^4}\right)+ {c_4 \over m^2} + \cdots,
\eqa
\label{Expanded:eN:amplitude:full:NREFT}
\eseq
where the ellipsis indicates terms proportional to powers of ${\bf p}$ and ${\bf p}^\prime$.
Both integrals in \eqref{Expanded:eN:amplitude:full:NREFT} are scaleless, thus vanish in DR.
A close look indicate that each of the integrals is severely IR divergent. However,
as warranted by the guiding principle of EFT, they must share the identical IR divergence.
Moreover, since these two expanded integrals turn out to be exactly identical at order $Z^2\alpha^2$,
there is no room for $c_4$ to contribute at this order.
Thus we have
\beq
c_4= 0 + \mathcal{O}(Z^3\alpha^3).
\label{c4:order:Z2alpha2}
\eeq
One may wonder that the vanishing of $c_4$ may persist to even higher order~\cite{Weinberg:1995mt}.
Nevertheless we will not dwell on this issue further in this work~\footnote{Recall that the famous
$\delta^{(3)}({\bf r})$-type Uehling potential, which arises from the vacuum polarization correction
to the photon propagator, may mimic the $c_4$ contact interaction.
Nevertheless, the vacuum polarization is a genuinely QED effect, which is not contained in the
Klein-Gordon equation \eqref{KG:Equation} at all. Therefore we discard this complication.}.

\section{Renormalization of Local composite operator $\varphi N$}
\label{Renorm:local:op:varphiN}

In our preceding work on the nonrelativistic Coulomb-Schr\"odinger EFT,
the local composite operator made of the electron and nucleus fields is free from
UV divergences~\cite{Huang:2018yyf}.
Nevertheless, in the enlarged NREFT defined in \eqref{NREFT:spin-0-electron:atom},
when a certain class of relativistic corrections are included,
it is conceivable that the composite operator $\varphi N$ becomes UV divergent,
and calls for renormalization.

The major goal of this section is pursue that, to which order in $Z\alpha$,
the {\it logarithmic} UV divergence would arise to necessitate the renormalization of the the $S$-wave composite
operator $\varphi N$. In some sense, this section is analogous to section~\ref{Renorm:Local:Composite:OP:sQED},
yet with the relativistic QFT of \eqref{UV:atom:lagrangian}
replaced with the NREFT of \eqref{NREFT:spin-0-electron:atom}.

As inspired by the analysis based on the quantum-mechanical perturbation theory
in section~\ref{Quantum:perturb:corr:wave:func:origin}, we see that the ${\bf p}^4$ kinetic correction
does bring the logarithmic UV divergence, while the Darwin term does not.
Therefore, in our explicit calculation, we will not include ${\bf p}^6$ kinetic correction and the Darwin term.

Similar to \eqref{Renormalize:local:composite:op}, we adopt a multiplicative renormalization procedure for
the local operator $\varphi N$:
\beq
[\varphi N]_R = Z_{\mathcal S}\, \varphi_R N_R = Z_{\varphi N}\, \varphi^{(0)} N^{(0)},
\label{Renormalize:composite:op:for:NREFT}
\eeq
where $Z_\mathcal{S}$ and $Z_{\varphi N}$ are  divergent renormalization constants.
Due to the single-pole structure of the nonrelativistic propagators
and the instantaneous Coulomb interaction, self-energy diagrams for the $\varphi$ and $N$ fields
simply vanish in arbitrarily high order.
As a consequence, we do not bother to distinguish the bare and renormalized fields for $e$ and $N$,
therefore, $Z_\mathcal{S}=Z_{\varphi N}$ in \eqref{Renormalize:composite:op:for:NREFT}.

\begin{figure}[tb]
\centering
\includegraphics[width=1.0\textwidth]{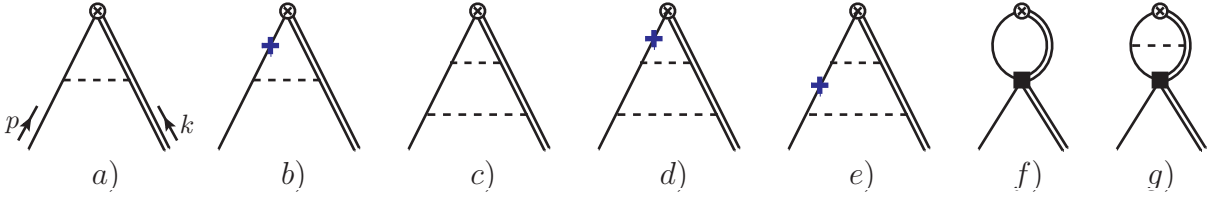}
\caption{Some representative higher-order NREFT diagrams for ${\mathcal G}({\bf{p}},E)$ defined in
\eqref{Three-point:Green:function:NREFT}. Except for the last two diagrams that
contain a contact interaction vertex, all other diagrams arise at most at ${\cal O}(Z^2\alpha^2)$,
as well as contain at most a single insertion of the ${\bf p}^4$ kinematic term.
\label{Fig:vertex:corr:all}}
\end{figure}

In analogue with \eqref{Three-point:Green:function},
for future use we also introduce the three-point Green function
inserted with the renormalized local operator $[\varphi N]_R$:
\beq
{\mathcal G}({\bf{p}},E\equiv p^0+k^0) \equiv \int \! d^4y \, d^4z\  e^{{\color{red} -} ip\cdot y {\color{red} -} ik\cdot z}
\langle 0 |T \left\{ [\varphi N]_R(0) \varphi^\dagger (y) N^\dagger(z) \right\}|0\rangle_{\rm amp}.
\label{Three-point:Green:function:NREFT}
\eeq
The subscript ``amp'' again implies that the external electron and nucleus legs get amputated.
Some higher-order corrections to this Green function through ${\cal O}(Z^2\alpha^2)$
are portrayed in Fig.~\ref{Fig:vertex:corr:all}.
We assume the external momenta ${\bf p}$ and $k$ to be soft, in the sense that they are of order $mZ\alpha \ll m$.
For simplicity, we further assume $E\equiv p^0+k^0<0$, and define $\kappa=\sqrt{-2m E}>0$.

The NLO correction depicted in Fig.~\ref{Fig:vertex:corr:all}$a)$ is finite,
\bqa
{\mathcal G}^{(1)}_{\ref{Fig:vertex:corr:all}a} ({\bf{p}},E) &=&
Ze^2 \mu^{\epsilon} \int \!\! \frac{d^Dl}{(2\pi)^D} {\cal D}_e(l)D_{00}(p-l)D_N(k+p-l)
\nn\\
&= & 2mZe^2\int \!\! \frac{d^3{\bf l}}{(2\pi)^3}\frac{1}{({\bf l}^2+\kappa^2)({\bf l}-{\bf p})^2}
=\frac{2mZ\alpha}{ |{\bf p}|} \tan^{-1}\!{|{\bf p}|\over \kappa},
\label{Fig7a:two:loop:vertex}
\eqa
hereafter we use the superscript $(n)$ to denote the order of $Z\alpha$
for the corresponding diagram.

With one insertion of the kinetic correction onto the electron propagator,
Fig.~\ref{Fig:vertex:corr:all}$b)$ turns to be linearly UV divergent.
Nevertheless, it gives finite result within DR,
\bqa
\label{Fig7b:two:loop:vertex}
{\mathcal G}^{(1)}_{\ref{Fig:vertex:corr:all}b} ({\bf{p}},E) &=&
i\mu^{ \epsilon}Ze^2\int \!\! \frac{d^D l}{(2\pi)^D} {\cal D}_e^2(l)D_{00}(p-l)D_N(k+p-l) {{\bf l^4}\over 8m^3}
\\
&=& -{Ze^2\over 2m}\int\!\! \frac{d^3{\bf l}}{(2\pi)^3}\frac{{\bf l}^4}{({\bf l}^2+\kappa^2)^2({\bf l}-{\bf p})^2}
=\frac{2E Z\alpha}{|{\bf p}|}\left(\tan^{-1}\! {|{\bf p}|\over \kappa}-
\frac{\kappa |{\bf p}|}{4({\bf p}^2+\kappa^2)}\right),
\nn
\eqa
which is order-$v^2$ suppressed relative to \eqref{Fig7a:two:loop:vertex}.
Furthermore, one can prove that, the one-loop vertex diagram with arbitrary number of insertion of
the kinetic corrections would never bring logarithmic divergence in DR.
The technical reason is because the argument of the Gamma function is always
half of an odd integer when $D\to 4$.

At next-to-next-to-leading order (NNLO), the two-loop diagram without insertion of relativistic correction, as depicted in
Fig.~\ref{Fig:vertex:corr:all}$c)$, is still UV finite.
Depending on which electron propagator is inserted with the kinetic term, we are left with two
distinct diagrams, Fig.~\ref{Fig:vertex:corr:all}$d)$ and Fig.~\ref{Fig:vertex:corr:all}$e)$, respectively.

We first consider the relativistic correction is inserted in the upper loop, as
indicated in Fig.~\ref{Fig:vertex:corr:all}$d)$. Some straightforward
calculation leads to
\bqa
\label{Fig7d:two:loop:vertex}
&& {\mathcal G}^{(2)}_{\ref{Fig:vertex:corr:all}d}({\bf{p}},E) =
 iZ^2e^4\mu^{2\epsilon}\int \!\! \frac{d^D l }{(2\pi)^D}\int\!\! \frac{d^D q}{(2\pi)^D}
 {\cal D}_e(l )D_{00}(p-l )D_N(k+p-l )
\nn\\
& \times & {\cal D}_e^2(q)D_{00}(l -q) D_N(k+p-q) {{\bf q}^4\over 8m^3}
\\
& = &Z^2e^4 \mu^{2\epsilon}
\int\!\! {d^{D-1}{\bf l}  \over (2\pi)^{D-1}} \int \!\! {d^{D-1}
{\bf q} \over (2\pi)^{D-1}} {1\over ({\bf l} -p)^2}  {1\over ({\bf l} -{\bf q} )^2} {1\over {\bf l}^2+\kappa^2}
{{\bf q}^4 \over ({\bf q}^2+\kappa^2)^2}
\nn\\
& = &
	Z^2e^4\mu^{2\epsilon}
\int\!\! {d^{D-1}{\bf l}  \over (2\pi)^{D-1}} \int \!\! {d^{D-1}
{\bf q} \over (2\pi)^{D-1}}
{1\over ({\bf l} -p)^2} {1\over ({\bf l} -{\bf q} )^2} {1\over {\bf l}^2+\kappa^2}\left[1- {2\kappa^2\over {\bf q}^2+\kappa^2}+ {\kappa^4\over ({\bf q}^2+\kappa^2)^2}\right].
\nn
\eqa
Simple counting of the superficial degrees of divergence implies that,
in order to nail down the potential UV divergence, it is sufficient to retain only
the first term in the bracket. For the integral over ${\bf q}$,
one may simply shift the variable and obtain
\beq
\int\!\! {d^{D-1}{\bf{q}}\over (2\pi)^{D-1}} {1\over {\bf q}^2}=0,\qquad\qquad {\rm DR}
\eeq
since scaleless integrals are set to zero in dimensional regularization.
Obviously, this integral actually corresponds to a linear UV divergence
provided the momentum hard cutoff is imposed.
It is reminiscent of the leading linear divergence encountered in \eqref{Kin:corr:wf},
the first-order correction to the Schr\"{o}dinger wave function at the origin from the
${\bf p}^4$ kinetic perturbation.

So far, the diagrams we encountered are either UV finite or power divergent.
In fact, Fig.~\ref{Fig:vertex:corr:all}$e)$,
the NNLO diagram with the kinetic term inserted on the
electron propagator in the lower loop, is the most interesting one,
because it gives the desired logarithmic singularity~\footnote{Note this kind of  
logarithmic UV singularity at two loop was first discovered when 
matching vector current from QCD onto NRQCD~\cite{Czarnecki:1997vz,Beneke:1997jm}. For
a calculation directly from the nonrelativistic EFT, see also \cite{Luke:1999kz}.}:
\bqa
\label{Fig7e:two:loop:vertex}
& & {\mathcal G}^{(2)}_{\ref{Fig:vertex:corr:all}e}({\bf p},E) =
iZ^2e^4\mu^{2\epsilon}
\int\!\! {d^D l  \over (2\pi)^D} \int\!\!
{d^D q \over (2\pi)^D} {\cal D}_e^2(l )D_{00}(p-l )D_N(k+p-l )
\nn\\
&\times &  {\cal D}_e(q) D_{00}(l -q) D_N(k+p-q)
{{\bf l}^4\over 8m^3}
\nn\\
&\approx &  Z^2 e^4 \mu^{2\epsilon} \int\!\!
{d^{D-1}{\bf l} \over (2\pi)^{D-1}}\int\!\! {d^{D-1}{\bf q}  \over (2\pi)^{D-1}}
{1\over {\bf q}^2{\bf l}^2({\bf l} -{\bf q} )^2}\bigg|_{{\bf q},\:{\bf l}\;{\rm hard}}
\\
&=& Z^2\alpha^2\left({1\over 2\epsilon}+\ln \mu\right)+\text{finite terms}.
\nn
\eqa
Note that the UV divergence comes from the region when both loop momenta ${\bf q}$
and $\bf l$ become hard ($\sim m$). Simple power counting indicates the above
two-loop integration does result in a logarithmic UV singularity.

Last but not the least, in principle one should also consider the one-loop diagram
involving a contact interaction vertex, as depicted in Fig.~\ref{Fig:vertex:corr:all}$f)$.
This diagram turns out to be linearly UV divergent, yet without subleading logarithmic UV divergence.
On the contrary, the two-loop diagram involving the contact interaction,
Fig.~\ref{Fig:vertex:corr:all}$g)$, does yield a leading logarithmic UV divergence.
In accordance with \eqref{c4:order:Z2alpha2}, $c_4$ is at least of order $Z^3\alpha^3$,
therefore Fig.~\ref{Fig:vertex:corr:all}$g)$ would contribute a logarithmic divergence
at least at order-$Z^4\alpha^4$, which is beyond the targeted accuracy of $Z^2\alpha^2$.

To summarize, among all the higher-order NREFT diagrams for ${\mathcal G}({\bf{p}},E)$
through the order-$Z^2\alpha^2$, only Fig.~\ref{Fig:vertex:corr:all}$e)$ contributes an
intended logarithmic UV divergence. The operator renormalization constant $Z_{\mathcal{S}}$ can thus be determined:
\beq
Z_{\mathcal{S}}=1-\frac{Z^2\alpha^2}{2\epsilon}+\cdots,
\label{compo-renorm}
\eeq
in the MS scheme.
The corresponding anomalous dimension of the composite operator $\varphi N$ is thus
\beq
\gamma_{\mathcal{S}}\equiv\frac{d\log{Z_\mathcal{S}}}{d\log\mu}= Z^2\alpha^2.
\label{ano-dim}
\eeq

\section{Operator Product Expansion in NREFT}
\label{sec:OPE:NREFT}

In section~\ref{sQED:OPE:failure}, we have considered the Wilson expansion of the electron field in scalar QED and the nucleus field
in HNET. The Wilson coefficient associated with the leading $S$-wave operator begins with $\ln r$ at first order in $Z\alpha$.
This incorrect coefficient of $Z\alpha$ contradicts that in the KG wave function near the origin, thus we have to give up the
OPE in the relativistic scalar QED. Numerous evidences have accumulated from section~\ref{Quantum:perturb:corr:wave:func:origin} and \ref{Renorm:local:op:varphiN}, which encouragingly suggest that the correct route
is to formulate the OPE in the context of nonrelativistic EFT,
by taking a certain class of relativistic corrections into account.
In this section, we will pursue this trail comprehensively.

We will consider the product of $\varphi$ and $N$ at equal time,
in the framework of the NREFT specified in \eqref{NREFT:spin-0-electron:atom}.
In the ${\bf r}\to {1\over m}$ limit~\footnote{In a renormalizable QFT,
such as the sQED considered in section~\ref{Renorm:local:op:varphiN},
the Wilson expansion can be conducted literally in the $r\to 0$ limit.
However, as a low-energy effective theory,
our NREFT remains only valid when the probed distance is greater than the electron's Compton wavelength.
Therefore, all the OPE relations in this section should be understood to work only
in the limit $r\to {1\over m}$.}, the Wilson expansion is anticipated to read~\footnote{Note for the electron field appearing in OPE, 
we choose the nonrelativistic NRQED field in the spirit of 
\cite{Caswell:1985ui,Bodwin:1994jh}, rather than the lower energy EFT such as the 
potential NRQED~\cite{Brambilla:2004jw}, since we need to work with a local EFT.}
\bseq
\bqa
\varphi({\bf r}) N(0) &=&  {\mathcal C}(r) [\varphi N]_R(0)+\cdots,
\\
{\mathcal C}(r) &=& 1+ c^{(1)}(r) +  c^{(2)}(r)+\cdots,
\eqa
\label{OPE:NREFT:coordinate:space}
\eseq
with $r\equiv |{\bf r}|$. Hereafter the superscript $(n)$ again indicates the power of $Z\alpha$.

It is also useful to express the OPE in momentum space:
\bseq
\bqa
\widetilde{\varphi}({\bf q}) N(0) & \equiv &
\int\!\! d^3{\bf{r}}\, e^{-i{\bf{q}}\cdot {\bf{r}}}\varphi({\bf r}) N(0) =
\widetilde{\cal C}(q)[\varphi N]_R(0)+\cdots,
\\
\widetilde{\mathcal C}(q) &=& (2\pi)^3\delta^{(3)}({\bf q}) + \tilde{c}^{(1)}(q) +  \tilde{c}^{(2)}(q)+\cdots,
\eqa
\label{OPE:NREFT:momentum:space}
\eseq
with $q\equiv |{\bf q}|\to m$.

Very recently we have determined $c^{(1)}(r)=-mZ\alpha$ and $\tilde{c}^{(1)}(q) =\frac{8\pi mZ\alpha }{{\bf q}^4}$
in the nonrelativistic  EFT~\cite{Huang:2018yyf}.
In the rest of the section, we will add the ${\bf p}^4$ kinetic correction into the Coulomb-Schr\"{o}dinger EFT,
and verify the OPE relations \eqref{OPE:NREFT:coordinate:space} and \eqref{OPE:NREFT:momentum:space}
through the order $Z^2\alpha^2$, in both momentum space and coordinate space.
The direct outcome is to ascertain the Wilson coefficients $c^{(2)}(r)$ and $\tilde{c}^{(2)}(q)$.
We will verify that $c^{(2)}(r)$ captures the desired $Z^2\alpha^2\ln r$ behavior,
in perfect agreement with the near-the-origin behavior of the $S$-wave KG wave function.
In the end we will employ the renormalization group equation (RGE) to resum large logarithms of
$r/a_0$ to all orders.

\subsection{OPE in Momentum Space}
\label{subsec:OPE:MOM}

We begin with examining the OPE relation \eqref{OPE:NREFT:momentum:space} in the momentum space.
To facilitate the study, in analogue with \eqref{Four-point:Green:function:momentum},
we introduce the following amputated four-point Green function:
\beq
\widetilde{\Gamma}\left({\bf{q}};{\bf{p}},E\equiv k^0+p^0\right)
\equiv \int\!\! d^4y \, d^4z\,
e^{-ip\cdot y-ik\cdot z} \langle 0 |T \left\{ \widetilde{\varphi}({\bf q}) N(0)
\varphi^\dagger (y) N^\dagger (z)\right\}|0\rangle_{\rm amp}.
\label{Four:point:Green:function:momentum:space:NREFT}
\eeq
As before, the  momentum ${\bf q}$ is hard ($\sim m$), which flows into the upper-left electron line
and exits from the upper-right nucleus line, while the momenta $p$ and $k$ signify the
soft momenta carried by the amputated electron and nucleus, with $p, k\ll m$.
The key is to verify that $\widetilde{\Gamma}$ does possess the following factorized form,
order by order in $Z\alpha$:
\beq
\widetilde{\Gamma}({\bf{q}};p,k) \xrightarrow{{\bf q} \to m} \widetilde{\cal C}(q)\,
{\cal G}({\bf p}, E)+\cdots,
\label{OPE:NREFT:momentum:space:verification}
\eeq
where the three-point Green function with a $\varphi N$ insertion
has been defined in \eqref{Three-point:Green:function:NREFT}.

To examine \eqref{OPE:NREFT:momentum:space:verification} in perturbation theory,
we write $\widetilde{\Gamma}=\sum_{n=1}^\infty \widetilde{\Gamma}^{(n)}$,
where the superscript $n$ denotes the order in $Z\alpha$.
Thus at a given order $n$, \eqref{OPE:NREFT:momentum:space:verification} becomes
\beq
\widetilde{\Gamma}^{(n)}({\bf{q}};{\bf{p}},E)\xrightarrow{{\bf q} \to m}
\sum_{i=1}^n \tilde{c}^{(i)}(q) \,{\cal G}^{(n-i)}({\bf p},E)+\cdots.
\label{OPE:mom:space:finite:order}
\eeq
It is worth mentioning that, whereas $\tilde{c}^{(n)}(q)$ has to begin from $n=1$,
${\cal G}^{(n)}$ actually starts from $n=0$.

\subsubsection{$\bf q$-scaling of $\widetilde{\Gamma}$ and our selecting recipe}

\begin{figure}[tb]
\centering
\includegraphics[width=0.9\textwidth]{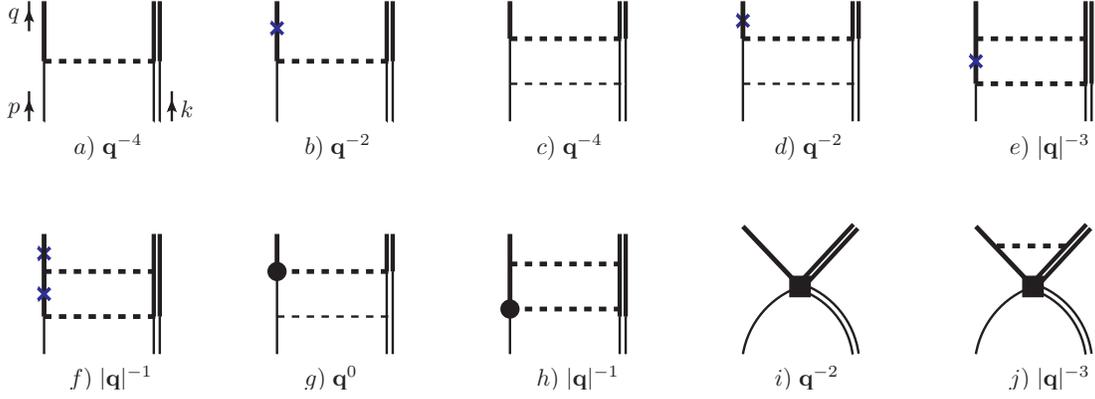}
\caption{Leading ${\bf q}$-scaling behavior of numerous higher-order diagrams
for the momentum-space Green function $\widetilde{\Gamma}$.
The thick lines are meant to carry the hard momentum of order ${\bf q}$,
in order to contribute to the specified leading region.
The cross, heavy dot, solid square refer to the ${\bf p}^4$ kinetic term,
Darwin term and contact interaction, respectively.
\label{fig:qscaling}}
\end{figure}

Before moving into the detailed calculation, it is instructive to outline
the leading scaling behavior with respect to the hard momentum ${\bf q}$ for
a general four-point diagram.
Recall that in \cite{Huang:2018yyf}, we proved that all the diagrams in
Coulomb-Schr\"{o}dinger theory exhibit a universal leading scaling behavior $\propto {\bf q}^{-4}$.
Nevertheless, with the relativistic correction included in the NREFT,
it seems that we have opened Pandora's box -- in the sense that arbitrary scaling
$\propto {\bf q}^{n}$ with $n$ an integer greater than $-4$ might arise.
In Fig.~\ref{fig:qscaling}, we have shown some characteristic tree and one-loop diagrams,
whose leading asymptotic behaviors may range from ${\bf q}^{-4}$ to ${\bf q}^0$.

Were $\tilde{c}(q)$ legally containing ${\bf q}^{-2}$ (Fig.~\ref{fig:qscaling}$b$, $d$, $i$),
$\vert {\bf q}\vert^{-1}$ (Fig.~\ref{fig:qscaling}$f$, $h$) and ${\bf q}^{0}$ (Fig.~\ref{fig:qscaling}$g$),
the corresponding Wilson coefficient in coordinate space would be proportional to
$1/r^2$, $1/r$ and $\delta^{(3)}({\bf r})$, respectively.
These near-the-origin singular behaviors are much more severe than the
logarithmic singularity exhibited by the actual KG wave function,
and are not even square integrable.

It may seem natural to speculate whether there is any objective criterion to determine
which types of diagrams we should keep or not?
We first note that, when pinching the external electron and nucleus
lines from the top of the diagrams in Fig.~\ref{fig:qscaling},
one then obtains the three-point Green function ${\cal G}$
defined in \eqref{Three-point:Green:function:NREFT}, some of which have been portrayed
in Fig.~\ref{Fig:vertex:corr:all}.
It is easy to see that those  ${\bf q}^{n}$ ($n>-4$) leading scaling of
the four-point diagrams would correspond to the $\Lambda^{n+3}$ UV power divergences in ${\cal G}$.

In section~\ref{Quantum:perturb:corr:wave:func:origin} about quantum mechanic perturbation theory,
we have seen the emergence of various UV power divergences in \eqref{UV:Div:Corrs:to:WF:DeltaH},
which are nevertheless absent in the original KG wave function \eqref{KG:WF:origin:Log:Div}.
There we argued that since the hard momentum cutoff $\Lambda$ violates the spacetime symmetry,
it may lead to some artificial and unphysical effects. Dimensional regularization turns out to be
much more advantageous, not merely convenient.
It is well-known that DR can only access the logarithmic rather than power divergences.
Therefore we have agreed in section~\ref{Renorm:local:op:varphiN}
to exclusively use DR as the UV regulator.
Since the Wilson coefficient in OPE is intimately connected with the renormalization of local composite operator,
it may sound reasonable that in our case, we can throw away all the contributions to the Green function $\widetilde{\Gamma}$
with a scaling behavior $\propto |{\bf q}|^{n}$ with $n > -3$.

Another reason is from practical consideration.
As indicated in \eqref{KG:WF:origin:Log:Div},
the worst near-the-origin behavior of the KG wave function for $S$-wave hydrogen-like atom can only
be logarithmic. To be capable for accounting for this universal short-range behavior,
we should establish some organizing principles, which justify abandoning
all the contributions to $\widetilde{\Gamma}$ that has a scaling behavior
$\propto |{\bf q}|^{n}$ with $n \ge -2$.

Therefore, let us state explicitly our recipe of analyzing the asymptotic behavior of
momentum-space four-point function $\widetilde{\Gamma}$.
In the $mZ\alpha\ll {\bf q}\sim m$ limit, to ascertain the Wilson coefficients affiliated with the
local composite operator $[\varphi N]_R$,  we only need single out those contributions
to the Green function $\widetilde{\Gamma}$ with the $1/|{\bf q}|^3$ and $1/{\bf q}^4$ scaling:
\beq
\widetilde{\Gamma}({\bf{q}};\,{\bf{p}},E) \xrightarrow{q\to m}
{1\over {\bf q}^4}\quad{\rm and}\quad {1\over \vert{\bf q}\vert^3},
\label{selecting:strategy}
\eeq
with the accompanying factors can only involve $m$ and $Z\alpha$, instead of $\bf p$ and $\kappa$.
As will be made clear later, this specific selecting recipe would lead to
the so-called ``double-layer form of the OPE''.

A quantification for \eqref{selecting:strategy} is in order. By retaining only the ${\bf q}^{-4}$ or
$|{\bf q}|^{-3}$ terms, we do not mean that we should neglect those diagrams with the leading scaling behavior
$\propto |{\bf q}|^{n}$ ($n \ge -2$ as a whole, such as Fig.~\ref{fig:qscaling}$b)$, $d)$.
Upon the expansion in $1/|{\bf q}|$, as long as these diagrams would contain the ${\bf q}^{-4}$ or
$|{\bf q}|^{-3}$ terms in the subleading order, there would be no reason to reject such contributions
in our analysis.

It is not straightforward to present an all-order proof for the OPE identity in momentum space
\eqref{OPE:NREFT:momentum:space:verification}. Nevertheless, we will be satisfied
if the OPE pattern can be nontrivially uncovered in the first few orders in $Z\alpha$, in the
spirit of \eqref{OPE:mom:space:finite:order}.
Concretely speaking, in the remaining subsections,
we will investigate the asymptotic behavior of the diagrams in Fig.~\ref{fig:qscaling},
verify the OPE indeed holds through order $Z^2\alpha^2$,
and also determine the corresponding Wilson coefficients.

\begin{figure}[tb]
\centering
\includegraphics[width=0.9\textwidth]{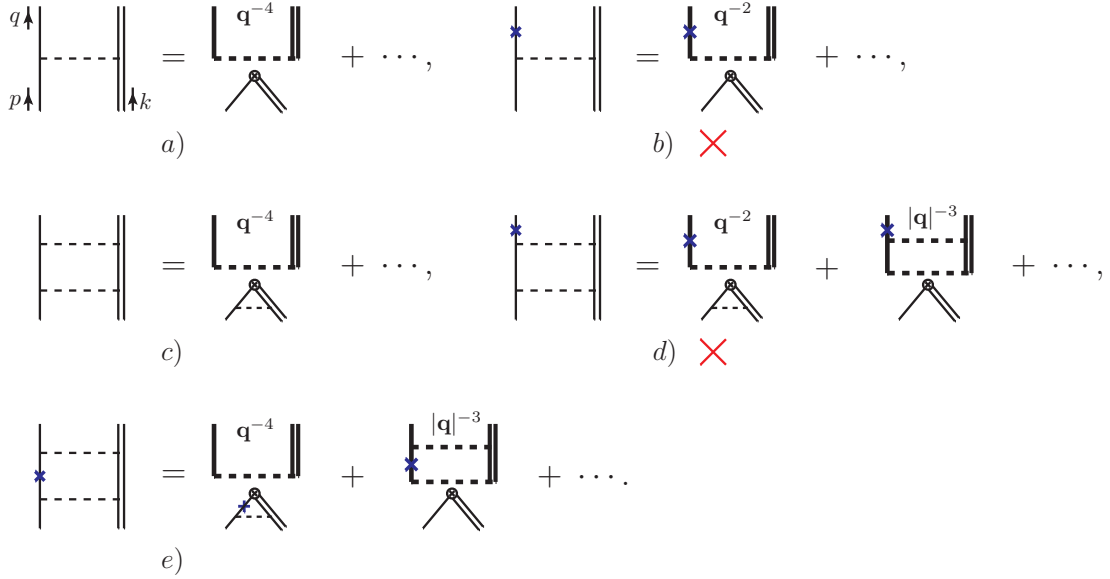}
\caption{The factorization of various four-point Green functions in momentum space
into hard and soft parts through order $Z^2\alpha^2$. The red cross implies that the corresponding
diagram can be discarded when considering its contribution to the OPE, in light of
\eqref{selecting:strategy}.
\label{Factorization:momentum:space}}
\end{figure}

In Fig.~\ref{Factorization:momentum:space}, we present a closer examination on the
asymptotic scaling of those most relevant diagrams in Fig.~\ref{fig:qscaling},
by explicitly exhibiting their factorized structures in the large-${\bf q}$ limit.

\subsubsection{Determining $\tilde{c}^{(1)}(q)$}

Let us first consider Fig.~\ref{Factorization:momentum:space}$a)$, the simplest tree diagram at order $Z\alpha$,
which was already analyzed in \cite{Huang:2018yyf}:
\bqa
\widetilde{\Gamma}^{(1)}_{\ref{Factorization:momentum:space}a}
({\bf{q}};\,{\bf{p}},E) &=& Ze^2\int\!\! \frac{d q^0}{2\pi} {\cal D}_e(q)D_{00}(p-q)D_N(k+p-q)
\nn\\
& = & - Ze^2 (2m)\frac{i}{{\bf q}^2-2mE}\frac{i }{({\bf q}-{\bf p})^2}
\nn\\
&\xrightarrow{q\to m}&  {8\pi mZ\alpha \over {\bf q}^4}+\cdots.
\label{9a:LO:tree:factorization:mom:space}
\eqa
In the last line, we have used the fact $|{\bf q}|\gg |{\bf p}|, E$.
In light of \eqref{selecting:strategy}, we need only keep the leading term in the expansion.

As indicated by the in Fig.~\ref{Factorization:momentum:space}$a)$, the leading ${\bf q}^{-4}$ scaling
comes from the subdiagram where the hard momentum of order $|{\bf q}|$ flows into the upper-left electron line, passes
through the Coulomb ladder, then exits from the upper-right nucleus line. For clarity, we have represented the lines
carrying hard momentum by the thick lines.
The factorization of Fig.~\ref{Factorization:momentum:space}$a)$ into the product of the hard factor and soft
factor becomes manifest. In accordance with \eqref{OPE:mom:space:finite:order},
one readily reads off the Wilson coefficient from \eqref{9a:LO:tree:factorization:mom:space}:
\beq
\tilde c^{(1)}(q) = {8\pi mZ\alpha \over {\bf q}^4},
\label{tilde:c1:expression}
\eeq
where ${\cal G}^{(0)}=1$ has been used.

Next let us turn to Fig.~\ref{Factorization:momentum:space}$b)$, the tree-level diagram with one
insertion of the ${\bf p}^4$ kinetic term onto the inflowing electron propagator:
\bqa
\widetilde{\Gamma}^{(1)}_{\ref{Factorization:momentum:space}b}
({\bf{q}};\,{\bf{p}},E) &=& Ze^2\int\!\! \frac{d q^0}{2\pi}
{\cal D}^2_e(q){i {\bf q}^4\over 8m^3} D_{00}(p-q)D_N(k+p-q)
\nn\\
& = &  {Ze^2\over 2m} {{\bf q}^4 \over ({\bf q}^2+\kappa^2)^2 ({\bf q}-{\bf p})^2 }
\nn\\
&\xrightarrow{q\to m}&  {2\pi Z\alpha\over m {\bf q}^2} + \cdots.
\label{9b:LO:tree:factorization:mom:space}
\eqa
According to \eqref{selecting:strategy}, the leading $1/{\bf q}^2$ term should be dismissed.
The next term in the series expansion of \eqref{9b:LO:tree:factorization:mom:space}
is of the order $1/{\bf q}^4$, but is accompanied with the prefactor involving ${\bf p}$ or $\kappa$.
Therefore, these power-suppressed terms correspond to some operators other than
$\varphi N$, which can be safely neglected for our purpose.

By the similar analysis, adding more ${\bf p}^4$ kinematic corrections to the inflowing electron line,
or including higher-order relativistic corrections, would not induce new contribution to the
Wilson coefficient $\tilde c^{(1)}(q)$ as determined in \eqref{tilde:c1:expression}.

\subsubsection{Determining $\tilde{c}^{(2)}(q)$}

At order $Z^2\alpha^2$, one encounters the one-loop box diagrams as exemplified in
Fig.~\ref{fig:qscaling}$c)-h)$.
The last three diagrams, {\it e.g.} Fig.~\ref{fig:qscaling}$f)-h)$,
which contain the relativistic correction of $v^4/c^4$, exhibit the leading scaling behaviors of
$|{\bf q}|^{0}$ or $|{\bf q}|^{-1}$. One can verify that the sub-leading contributions to these diagrams
neither bear the desired $|{\bf q}|^{-3}$ or $|{\bf q}|^{-4}$ scaling, nor
generate new corrections to \eqref{tilde:c1:expression} for $\tilde c^{(1)}(q)$.
We thereby neglect these diagrams, and instead focus on Fig.~\ref{fig:qscaling}$c)-e)$, or equivalently, Fig.~\ref{Factorization:momentum:space}$c)-e)$.

It is possible to analytically work out these one-loop box integrals, but the resulting expressions
are rather cumbersome. Fortunately, to deduce the asymptotic results in large ${\bf q}$ limit,
it is more effective to take a shortcut, by first expanding the integrand prior to conducting the loop integration.
The loop momentum ${\bf l}$ can be classified into two categories: hard (${\bf l}\sim {\bf q}$) and
soft (${\bf l}\sim {\bf p},\kappa \ll {\bf q}$). This expansion-in-region technique
leads to readily manageable integrals, which result in compact expressions.

Let us first examine the box diagram without relativistic insertion, Fig.~\ref{Factorization:momentum:space}$c)$.
We can work out the integral as follows:
\bqa
\label{Fig:9c:mom:space:factorization}
\widetilde{\Gamma}^{(2)}_{\ref{Factorization:momentum:space}c}\left({\bf{q}};\,{\bf{p}},E\right) & =&
Z^2e^4 \mu^{2\epsilon}
\int\!\! {d q^0\over 2\pi}\, {\mathcal D}_e(q) D_N(k+p-q)
\nn\\
&\times &  \int\!\! {d^D l\over (2\pi)^D} {\cal D}_e(l) D_{00}(p-l) D_N(k+p-l) D_{00}(l-q)
\nn\\
& = & {4m^2 Z^2e^4\over {\bf q}^2+\kappa^2} \mu^{2\epsilon} \int\!\!
{d^{D-1}{\bf l}\over (2\pi)^{D-1}} {1\over ({\bf l}-{\bf q})^2 ({\bf l}-{\bf p})^2 \left({\bf l}^2+\kappa^2 \right)}
\\
&\approx & {4m^2 Z^2e^4\over {\bf q}^2} \left[ {1\over {\bf q}^2} \mu^{2\epsilon} \int_S
{d^{D-1}{\bf l}\over (2\pi)^{D-1}} {1\over ({\bf l}-{\bf p})^2 \left({\bf l}^2+\kappa^2 \right)}+
\mu^{2\epsilon} \int_H
{d^{D-1}{\bf l}\over (2\pi)^{D-1}} {1\over ({\bf l}-{\bf q})^2 {\bf l}^4}
\right]
\nn \\
&=& \tilde{c}^{(1)}(q)\,{\cal G}^{(1)}_{\ref{Fig:vertex:corr:all}a}({\bf p},E) + {\cal O}\left(1/|{\bf q}|^5\right),
\nn
\eqa
where $\tilde{c}^{(1)}(q)$ is given in \eqref{tilde:c1:expression}, the one-loop Green function ${\mathcal G}^{(1)}_{\ref{Fig:vertex:corr:all}a}$ is given in \eqref{Fig7a:two:loop:vertex}.
In the middle of derivation, we have explicitly decomposed the loop integral into the soft and hard parts,
which are labelled by the subscripts $S$ and $H$.
The contribution from hard loop region yields a subleading contribution of order $1/|{\bf q}|^5)$, thus has been neglected.
As can be visualized from the factorization structure in
Fig.~\ref{Factorization:momentum:space}$c)$,
this box diagram does not bring any new correction to ${\cal C}(q)$ at order $Z\alpha$~\cite{Huang:2018yyf}.

We then turn to Fig.~\ref{Factorization:momentum:space}$d)$, with a ${\bf p}^4$ kinetic term inserted onto
the inflowing electron line outside the box.  We have argued after \eqref{9b:LO:tree:factorization:mom:space}
that the analogous tree diagram in Fig.~\ref{Factorization:momentum:space}$d)$ needs not be considered.
Here let us examine this more complicated example, again employing the expansion-by-region technique:
\bqa
\label{Fig:9d:mom:space:factorization}
\widetilde{\Gamma}^{(2)}_{\ref{Factorization:momentum:space}d}({\bf{q}};\,{\bf{p}},E)
&=& Z^2e^4 \mu^{2\epsilon}
\int\!\! \frac{dq^0}{2\pi} \, {\cal D}^2_e(q)\, {i {\bf q}^4\over 8m^3}  D_N(k+p-q)
\nn\\
&\times & \int\!\! {d^Dl\over (2\pi)^D} {\cal D}_e(l) D_{00}(l-q) D_N(k+p-l) D_{00}(p-l)
\nn\\
&=& \frac{Z^2e^4{\bf q}^4}{({\bf q}^2+\kappa^2)^2} \mu^{2\epsilon} \int\!\!
\frac{d^{D-1}l}{(2\pi)^{D-1}} {1\over ({\bf l}-{\bf q})^2 ({\bf l}-{\bf p})^2 \left({\bf l}^2+\kappa^2 \right)}
\\
&\approx &  Z^2e^4 \left[ {1\over {\bf q}^2} \mu^{2\epsilon} \int_S
{d^{D-1}{\bf l}\over (2\pi)^{D-1}} {1\over ({\bf l}-{\bf p})^2 \left({\bf l}^2+\kappa^2 \right)}+
\mu^{2\epsilon} \int_H
{d^{D-1}{\bf l}\over (2\pi)^{D-1}} {1\over ({\bf l}-{\bf q})^2 {\bf l}^4}
\right]
\nn\\
&=& {2\pi Z\alpha\over m {\bf q}^2} \,{\cal G}^{(1)}_{\ref{Fig:vertex:corr:all}a}({\bf p},E) +
{2\pi^2 Z^2\alpha^2 \epsilon \over |{\bf q}|^3} \,{\cal G}^{(0)}({\bf p},E),
\nn
\eqa
where we have kept both contributions from the soft and hard regions,
with the factorization form illustrated in Fig.~\ref{Factorization:momentum:space}$d)$.
It is observed that the short-distance coefficient $\propto Z\alpha/m{\bf q}^2$,
which stems from the soft region, is identical to
\eqref{9b:LO:tree:factorization:mom:space} in Fig.~\ref{Factorization:momentum:space}$b)$.
There we have argued such term should be dismissed.
On the contrary, the contribution from the hard region appears to bear a factorized structure.
Nevertheless, the would-be Wilson coefficient $Z^2\alpha^2/|{\bf q}|^3$ has a prefactor $D-4$,
thus cannot contribute to the desired $Z^2\alpha^2\ln r$ term
upon Fourier transform~\footnote{This is compatible with the fact that
the two-loop vertex diagram with a ${\bf p}^4$ insertion in the
upper loop, Fig.~\ref{Fig:vertex:corr:all}$d)$, does not contain a
logarithmical UV divergence, as analysed in \eqref{Fig7d:two:loop:vertex}.}.
Therefore, we can discard the contribution of
Fig.~\ref{Factorization:momentum:space}$d)$ as a whole.

Our final stake is then in Fig.~\ref{Factorization:momentum:space}$e)$, which hopefully
serves the key to account for the $Z^2\alpha^2\ln r$ term in the KG wave function near the origin.
In fact, this expectation is well-grounded. As indicated in \eqref{Fig7e:two:loop:vertex},
the ${\cal O}(Z^2\alpha^2)$ vertex diagram with a ${\bf p}^4$ insertion in the
lower loop, Fig.~\ref{Fig:vertex:corr:all}$e)$, is indeed logarithmically UV divergent.
Actually, we have already remarked that Fig.~\ref{fig:qscaling}$e)$ has a
leading $1/|{\bf q}|^3$ scaling behavior when all the internal lines inside the loop become hard.
When sewing up two upper external lines in Fig.~\ref{fig:qscaling}$e)$,
one can recover the logarithmic UV divergence in Fig.~\ref{Fig:vertex:corr:all}$e)$.
As implied in Fig.~\ref{Factorization:momentum:space}$e)$, we can separate the contributions from
the soft and hard loop regions:
\bqa
\label{Fig:9e:mom:space:factorization}
\widetilde{\Gamma}^{(2)}_{\ref{Factorization:momentum:space}e} ({\bf{q}};\,{\bf{p}},E) &=&
Z^2 e^4 \mu^{2\epsilon} \int\!\! {d q^0 \over 2\pi} \,{\mathcal D}_e(q) \,
D_N(k+p-q)
\nn\\
&\times &  \int\!\! {d^D l \over (2\pi)^D} {\cal D}_e^2 (l) D_{00}(l-p) D_N(k+p-l) D_{00}(l-q) {i{\bf l}^4\over 8m^3}
\nn\\
&=& {Z^2 e^4 \mu^{2\epsilon}\over {\bf q}^2+\kappa^2} \int\!\! {d^{D-1}{\bf l}\over (2\pi)^{D-1}}
{{\bf l}^4\over ({\bf l}-{\bf p})^2 ({\bf l}-{\bf q})^2 ({\bf l}^2+\kappa^2)^2}
\\
&\approx & {Z^2 e^4\over {\bf q}^2} \left[ {1\over {\bf q}^2} \mu^{2\epsilon} \int_S
{d^{D-1}{\bf l}\over (2\pi)^{D-1}} {{\bf l}^4\over ({\bf l}-{\bf p})^2
({\bf l}^2+\kappa^2)^2}+
\mu^{2\epsilon} \int_H
{d^{D-1}{\bf l}\over (2\pi)^{D-1}} {1\over ({\bf l}-{\bf q})^2 {\bf l}^2}
\right]
\nn \\
&= & \tilde{c}^{(1)}(q)\, {\cal G}^{(1)}_{\ref{Fig:vertex:corr:all}b}({\bf p},E)+   \tilde{c}^{(2)}(q)\, {\cal G}^{(0)}({\bf p},E),
\nn
\eqa
where the first term encapsulates the soft loop contribution,
and the second term encodes the hard loop contribution.
Note the former is entirely absorbed in the correction to the three-point Green function,
${\cal G}^{(1)}_{\ref{Fig:vertex:corr:all}b}$ as given in \eqref{Fig7b:two:loop:vertex}.
In contrast, the latter brings a new correction for
the Wilson coefficient at order-$Z^2\alpha^2$:
\beq
\tilde{c}^{(2)}(q)= -16\pi^2\mu^{2\epsilon}Z^2\alpha^2 {2^{5-2 D} \pi ^{2-\frac{D}{2}}
\over \Gamma \left(\frac{D-2}{2}\right)\sin {\pi(D-1)\over 2} }\frac{1}{|{\bf q}|^{7-D}}
\xrightarrow{D\to 4} \frac{2\pi^2Z^2\alpha^2}{|{\bf q}|^3}.
\label{tilde:c2:q}
\eeq
This $|{\bf q}|^{D-7}$ scaling is just responsible for the logarithmic UV divergence
observed in \eqref{Fig7e:two:loop:vertex}, which represents the two-loop vertex
correction of Fig.~\ref{Fig:vertex:corr:all}$e)$.

As was stated earlier, at order $Z^2\alpha^2$, there is no need to consider
diagrams implementing $v^4/c^4$ or higher-order relativistic corrections,
{\it e.g.} Fig.~\ref{fig:qscaling}$f)-h)$, with two or more insertions of ${\bf p}^4$ term or a
Darwin term. This omission follows from the hypothesized working principle of double-layer OPE in \eqref{selecting:strategy},
because these diagrams do not admit the intended $1/|{\bf q}|^3$
or $1/{\bf q}^4$ scaling behaviors.

\subsubsection{Final expression of the OPE in momentum space}

Following the iterative pattern \eqref{OPE:mom:space:finite:order}, we have determined
the Wilson coefficients up to the order $Z^2\alpha^2$. Plugging \eqref{tilde:c1:expression} and \eqref{tilde:c2:q}
into \eqref{OPE:NREFT:momentum:space}, we find that
the OPE in momentum space explicitly reads
\bqa
\widetilde{\varphi}({\bf q}) N(0)   =
\left({8\pi mZ\alpha \over {\bf q}^4} + {2\pi^2 Z^2\alpha^2\over |{\bf q}|^3}+\cdots \right) \,[\varphi N]_R(0)+\cdots.
\label{OPE:mom:space:FINAL}
\eqa
The ellipsis inside the parenthesis
indicates those uncalculated higher-order Wilson coefficients such as $Z^3\alpha^3 \ln\! |{\bf q}|/{\bf q}^4$ and
$Z^4\alpha^4 \ln\!|{\bf q}|/|{\bf q}|^3$. Upon Fourier transform, these may correspond to the
coordinate-space Wilson coefficients of the form $Z^3\alpha^3 r \ln r$ and $Z^4\alpha^4 \ln^2 r$,
which are truly present in the near-the-origin behavior of the KG wave function
in \eqref{KG:Universal:Small:r}.

\subsection{OPE in coordinate space}
\label{subsec:OPE:COOR}

We now supplement Section~\ref{subsec:OPE:MOM} with the corresponding
OPE relation in the coordinate space, \eqref{OPE:NREFT:coordinate:space}.
This will provide the key element to explain the weakly divergent
near-the-origin behavior of the $S$-wave KG wave function.
Analogous to \eqref{Four-point:Green:function:coordinate},
for convenience here we introduce the following four-point Green function
in the coordinate space:
\beq
\Gamma({\bf r};{\bf p},E)\equiv\int \!\! d^4 y  d^4 z \,e^{-ip\cdot y}e^{-ik\cdot z}\langle 0 |T\left\{\varphi({\bf r}) N(0)\varphi^\dagger(y)N^\dagger(z)\right\}|0 \rangle_{\rm amp},
\label{Four:point:Green:function:coordinate:space:NREFT}
\eeq
which can be reached from \eqref{Four:point:Green:function:momentum:space:NREFT} by
\beq
\Gamma({\bf r};{\bf p},E) = \int\!\! {d^3{\bf{q}} \over (2\pi)^3}\, e^{i{\bf{q}}\cdot {\bf{r}}} \,\widetilde{\Gamma}({\bf{q}};{\bf p}, E).
\label{Gamma:Gammatilde:FT}
\eeq
As before, the Green function is amputated with the lower lines carrying soft
momenta $p$ and $k$.

Our goal is to establish that, in the small $r$ ($|{\bf r}| \to{1\over m}$) limit,
$\Gamma$ exhibits the following factorized form order by order in $Z\alpha$:
\beq
\Gamma({\bf r};{\bf p},E)  \xrightarrow{|{\bf r}| \to {1\over m}} {\cal C}(r)\,
{\cal G}({\bf p}, E)+\cdots,
\label{OPE:NREFT:coordinate:space:verification}
\eeq
where ${\cal G}$ was first introduced in \eqref{Three-point:Green:function:NREFT},
signifying the amputated three-point Green function inserted with the composite operator
$[\varphi N]_R$.

Following \eqref{OPE:NREFT:coordinate:space}, we can decompose the Wilson coefficient in power series
of $Z\alpha$:
\beq
{\mathcal C}(r) = c^{(0)} + c^{(1)}(r) +  c^{(2)}(r)+\cdots.
\label{Wilson:coeff:coord:sp:series:expansion}
\eeq

The lowest-order Wilson coefficient can be readily deduced in free theory.
Analogous to what is portrayed in Fig.~\ref{Fig:free:theory:ope}, the Green function $\Gamma$
possesses a disconnected topology, yields the very simple answer:
\beq
\Gamma^{(0)}({{\bf r}};{\bf p},E)= e^{i{\bf p} \cdot{\bf r}} = 1+ {\cal O}(p r).
\label{Gamma:LO:Prediction:NREFT}
\eeq
Since ${\cal G}^{(0)}=1$, it is straightforward to read off $c^{(0)}=1$.

Eq.~\eqref{OPE:NREFT:coordinate:space:verification} can be recast into a iterative form,
which is amenable to perturbation theory,
\beq
\Gamma^{(n)}({{\bf r}};\,{\bf{p}},E)  \xrightarrow{r \to {1\over m}} \sum_{i=0}^n c^{(i)} (r)
{\cal G}^{(n-i)}({\bf{p}},E)
\label{Gamma:n:iterative:product:NREFT}
\eeq
In contrast with its counterpart in the momentum space \eqref{OPE:mom:space:finite:order},
here both $c^{(i)}(q)$ and ${\cal G}^{(i)}$ start from $i=0$.

In \eqref{selecting:strategy}, we put forward the hypothesized principle of ``double-layer form of the OPE''
in the momentum space version of OPE.
Translated into the coordinate space, this principle indicates that
we only need sort out those contributions
to $\Gamma$ with the following scalings:
\beq
\Gamma({\bf{r}};\,{\bf{p}},E) \xrightarrow{r\to {1\over m}}
\ln^n \! r \quad{\rm and}\quad r \ln^n \! r. \qquad(n\ge 0)
\label{selecting:strategy:coor:space}
\eeq
It is understood that the accompanying factors can only be composed of $m$ and $Z\alpha$.

In the rest of this subsection, we will compute the higher-order Wilson coefficients through ${\mathcal O}(Z^2\alpha^2)$.

\subsubsection{Determining ${c}^{(1)}(r)$}
\label{subsubsec:OPE:COOR:c1}

\begin{figure}[tb]
\centering
\includegraphics[width=1.0\textwidth]{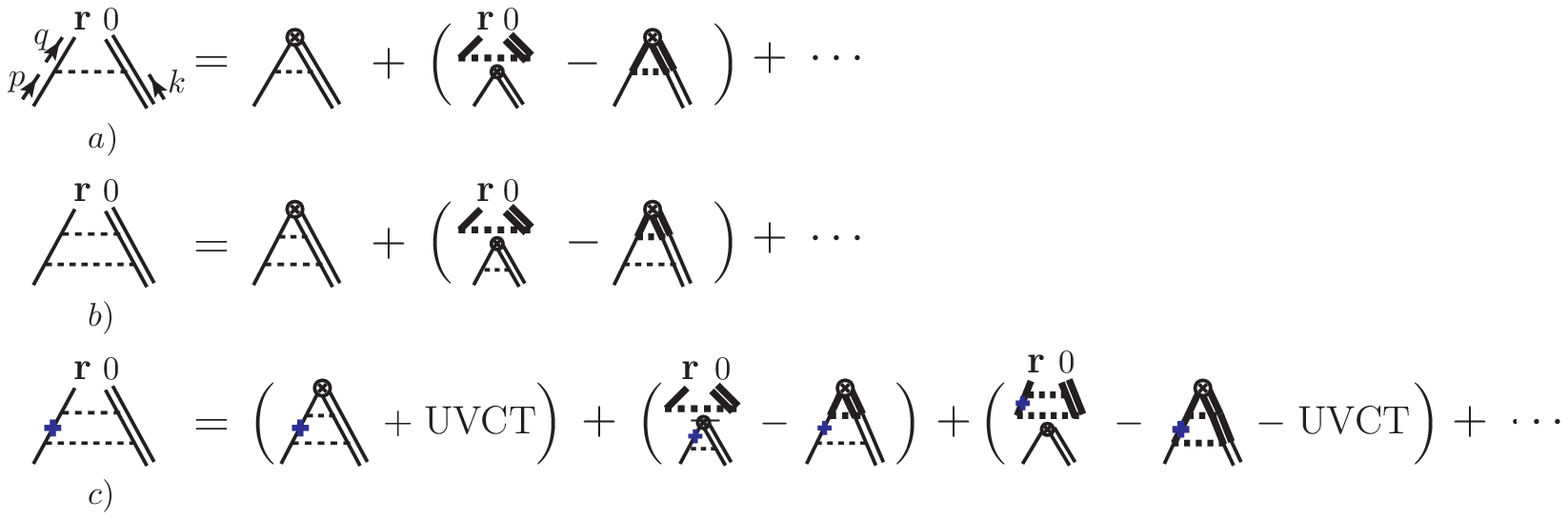}
\caption{Manifestation of the OPE structure in the coordinate-space Green function through order
$Z^2\alpha^2$. The Wilson coefficient $\sim Z^2\alpha^2\ln r$ is stemming from the parenthesis
in the last line.
\label{Wilson:coeff:coor:space:decomp}}
\end{figure}

Directly Fourier transforming $\tilde{c}_1(q)$ in \eqref{tilde:c1:expression}
to deduce ${c}^{(1)}(r)$ in the the coordinate space turns out to be
plagued with IR divergence associated with ${\bf q}\to 0$,
which is clearly unacceptable.
As illustrated in Fig.~\ref{Wilson:coeff:coor:space:decomp}$a)$,
the correct way to proceed is by expanding \eqref{Gamma:n:iterative:product:NREFT} through
order $Z\alpha$:
\beq
{\Gamma}_{\ref{Wilson:coeff:coor:space:decomp}a}^{(1)}({{\bf r}};\,{\bf{p}},E)
\xrightarrow{r\to {1\over m}} c^{(0)}(r)
{\cal G}_{\ref{Fig:vertex:corr:all}a}^{(1)}({\bf{p}},E)+ c^{(1)}(r) {\cal G}^{(0)}({\bf{p}},E),
\label{Gamma1:coor:space:decomposition}
\eeq
where the one-loop correction to the three-point Green function,
${\cal G}_{\ref{Fig:vertex:corr:all}a}^{(1)}$,
is given in \eqref{Fig7a:two:loop:vertex}.
Therefore, ${c}^{(1)}(r)$ can be obtained from the $\Gamma^{(1)}({\bf r})$ by subtracting off
the local Green function ${\cal G}_{\ref{Fig:vertex:corr:all}a}^{(1)}$:
\bqa
c^{(1)}(r) &=& \lim_{r\to {1\over m}} \Gamma_{\ref{Wilson:coeff:coor:space:decomp}a}^{(1)}({{\bf r}};\,{\bf{p}},E) - {\cal G}_{\ref{Fig:vertex:corr:all}a}^{(1)}({\bf{p}},E)
\nn\\
&=&  \int\!\!
{d^3{\bf q}\over (2\pi)^3} \widetilde{\Gamma}_{\ref{Wilson:coeff:coor:space:decomp}a}^{(1)}({\bf q};{\bf p}, E)
\left( e^{i{\bf{q}}\cdot {\bf r}}-1\right)\Big|_{{\bf q}\;{\rm hard}} \approx
\int\!\! {d^3{\bf q}\over (2\pi)^3} \tilde{c}^{(1)}(q)
\left( e^{i{\bf{q}}\cdot {\bf r}}-1\right)
\nn\\
& =& 8\pi mZ\alpha \int\!\! {d^3{\bf q}\over (2\pi)^3}
{e^{i{\bf{q}}\cdot {\bf r}}-1\over {\bf q}^4} = -m Z\alpha r.
\label{OPE:NREFT:c1r:extract}
\eqa
We have used the Fourier representation \eqref{Gamma:Gammatilde:FT} and
\beq
{\cal G}^{(n)}({\bf p},E)= \int\!\! {d^3 {\bf{q}} \over (2\pi)^3} \widetilde{\Gamma}^{(n)}({\bf{q}};{\bf p}, E),
\qquad{\rm for}\;n\ge 1.
\eeq
Furthermore, in the second line of \eqref{OPE:NREFT:c1r:extract},
we have used \eqref{9a:LO:tree:factorization:mom:space} and \eqref{tilde:c1:expression} to
keep only the leading asymptotic form of $\widetilde{\Gamma}^{(1)}$ in the
${\bf q}\to m$ limit, since Wilson coefficient stems from the
hard loop momentum region. Notice the subtraction term automatically guarantees
$c^{(1)}(r)$ to be IR finite, which is a consequence of $c^{(0)}(r)=1$.
This analytic expression of $c^{(1)}(r)$ was first obtained within the Coulomb-Schr\"{o}dinger EFT~\cite{Huang:2018yyf}.

\subsubsection{Deciphering ${c}^{(2)}(r)$}

From \eqref{Gamma:LO:Prediction:NREFT} and \eqref{OPE:NREFT:c1r:extract}, we have inferred that
$c^{(0)}$ and $c^{(1)}$ to be $1$ and  $-m Z\alpha r$, respectively. We proceed to deduce
the expression of $c^{(2)}(r)$,
In accordance with \eqref{Gamma:n:iterative:product:NREFT},
higher-order Wilson coefficients account for the difference between the full four-point Green function and the three-point
Green functions multiplied by lower-order Wilson coefficients. In particular, the difference solely comes from the hard loop region.
In the following, we will explicitly analyze the asymptotic behaviors of
$\Gamma^{(2)}({{\bf r}};\,{\bf{p}},E)$ in the limit $r\to {1\over m}$, by identifying the hard loop mentum and expanding the integrands
accordingly. This analysis would lead to the OPE structure as exhibited by
\eqref{Gamma:n:iterative:product:NREFT}, where the hard part renders the Wilson coefficients, and
the soft part is absorbed by the three-point Green function ${\cal G}^{(i)}({\bf p},E)$.

Using \eqref{Fig:9c:mom:space:factorization}, \eqref{Gamma:Gammatilde:FT} and
\eqref{Fig7a:two:loop:vertex}, we find the difference between
$\Gamma^{(2)}_{\ref{Wilson:coeff:coor:space:decomp}b}$ and ${\cal G}^{(2)}_{\ref{Fig:vertex:corr:all}a}$ to be
\bqa
\label{Diff:Gamma:10b:and:7a}
& & \Gamma^{(2)}_{\ref{Wilson:coeff:coor:space:decomp}b}\left({\bf{r}};\,{\bf{p}},E\right)-c^{(0)}(r) {\cal G}^{(2)}_{\ref{Fig:vertex:corr:all}a}({\bf p},E)= Z^2e^4 \mu^{2\epsilon}
\int\!\! {d^D q\over 2\pi}\, {\mathcal D}_e(q) D_N(k+p-q) \left(e^{i{\bf q}\cdot {\bf r}}-1\right)
\nn\\
& &\times\int\!\! {d^D l\over (2\pi)^D} {\cal D}_e(l) D_{00}(p-l) D_N(k+p-l) D_{00}(l-q)
\\
& & =  4m^2 Z^2e^4\mu^{2\epsilon}\int \frac{d^{D-1}{\bf q}}{(2\pi)^{D-1}}{1\over {\bf q}^2+\kappa^2}  \int\!\!
{d^{D-1}{\bf l}\over (2\pi)^{D-1}} {1\over ({\bf l}-{\bf q})^2 ({\bf l}-{\bf p})^2 \left({\bf l}^2+\kappa^2 \right)}.
\nn
\eqa

Eq.~\eqref{Diff:Gamma:10b:and:7a} in the limit $r\to {1\over m}$
is dominated by the situation when $\bf q$ becomes hard. More specifically,
one should consider two situations: the top loop
is hard and bottom loop is soft, as well as both loop momenta are hard,
\bqa
\label{Diff:Gamma:10b:and:7a:expand:q:hard}
& & \Gamma^{(2)}_{\ref{Wilson:coeff:coor:space:decomp}b}\left({\bf{r}};\,{\bf{p}},E\right)- {\cal G}^{(2)}_{\ref{Fig:vertex:corr:all}a}({\bf p},E) \approx \Bigg(2m  Z e^2\mu^{\epsilon}\int\frac{d^{D-1}{\bf q}}{(2\pi)^{D-1}}{e^{i{\bf q}\cdot {\bf r}}-1\over {\bf q}^4}\Bigg) {\cal G}^{(1)}_{\ref{Fig:vertex:corr:all}a}({\bf p},E) \bigg|_{{\bf q}\;{\rm hard},\;{\bf l}\;{\rm soft}}
\nn\\
&&+\Bigg( 4m^2 Z^2e^4\mu^{2\epsilon}\int\frac{d^{D-1}{\bf q}}{(2\pi)^{D-1}}\int
{d^{D-1}{\bf l}\over (2\pi)^{D-1}} {e^{i{\bf q}\cdot {\bf r}}-1\over ({\bf l}-{\bf q})^2 {\bf l}^4}\Bigg)
\bigg|_{{\bf q},\:{\bf l}\;{\rm hard}}
\nn\\
& &= c^{(1)}(r)\,{\cal G}^{(1)}_{\ref{Fig:vertex:corr:all}a}({\bf p},E) + {\cal O}\left(r^2\right),
\eqa
which only involves the known Wilson coefficient at order $Z\alpha$,
as is represented in the parenthesis of the second line in Fig.~\ref{Wilson:coeff:coor:space:decomp}$b)$.
The both-loop-hard regions yield a contribution subleading in $r$, thus can be neglected according to
our assumption \eqref{selecting:strategy:coor:space}. Note \eqref{Diff:Gamma:10b:and:7a:expand:q:hard}
is compatible with \eqref{Gamma:n:iterative:product:NREFT}, with the conclusion
$c^{(2)}(r)=0$ from this diagram.

Finally let us turn to Fig.~\ref{Wilson:coeff:coor:space:decomp}$c)$,
the most interesting order-$Z^2\alpha^2$ diagram where a ${\bf p}^4$
relativistic correction inserted on the lower loop.
With the aid of \eqref{Fig:9e:mom:space:factorization}, \eqref{Gamma:Gammatilde:FT} and
\eqref{Fig7e:two:loop:vertex}, we find the difference between
$\Gamma^{(2)}_{\ref{Wilson:coeff:coor:space:decomp}c}$ and ${\cal G}^{(2)}_{\ref{Fig:vertex:corr:all}e}$ to be
\bqa
\label{Diff:Gamma:10c:and:7e}
& & \Gamma^{(2)}_{\ref{Wilson:coeff:coor:space:decomp}c}({\bf{r}};{\bf{p}},E)-
c^{(0)}(r)\left({\cal G}^{(2)}_{\ref{Fig:vertex:corr:all}e}({\bf p},E)+ {\rm UVCT}\right)
\nn\\
& = & Z^2e^4\mu^{2\epsilon} \Bigg[\int\!\! {d^D q\over (2\pi)^D} {\cal D}_e(q)
D_N(k+p-q) \int\!\! {d^D{l}\over (2\pi)^D}\left(e^{i{\bf q}\cdot {\bf x}}-1\right)
\\
& \times & {\cal D}_e^2(l) D_{00}(p-l) D_N(k+p-l) D_{00}(l-q) \frac{i{\bf l}^4}{8m^3} \Bigg]-\delta Z_{\cal S} {\cal G}^{(0)}
\nn\\
& = & Z^2e^4\mu^{2\epsilon}\Bigg[\int\!\! {d^{D-1}{\bf q}\over (2\pi)^{D-1}} {e^{i{\bf q}\cdot {\bf x}}-1\over {\bf q}^2+\kappa^2}
\int\!\! {d^{D-1}{\bf l}\over (2\pi)^{D-1}}\frac{{\bf l}^4}{({\bf l}-{\bf p})^2({\bf l}-{\bf q})^2({\bf l}^2+\kappa^2)^2}\Bigg]-\delta Z_{\cal S},
\nn
\eqa
where $\delta Z_{\cal S}$ is the UV counterterm to renormalize the composite operator $\varphi N$.
Its value in the MS prescription is given in \eqref{compo-renorm}.

Similar to \eqref{Diff:Gamma:10b:and:7a},  in the limit $r \to {1\over m}$
\eqref{Diff:Gamma:10c:and:7e} is again dominated by the situation when $\bf q$ becomes hard.
Specifically, we consider two situations: the top loop
is hard and bottom loop is soft, as well as both loop momenta are hard,
\bqa
\label{Diff:Gamma:10c:and:7e:expand:q:hard}
& & \Gamma^{(2)}_{\ref{Wilson:coeff:coor:space:decomp}c}\left({\bf{r}};\,{\bf{p}},E\right)- {\cal G}^{(2)}_{\ref{Fig:vertex:corr:all}e\,R}({\bf p},E) \approx
\left(2m  Z e^2\mu^{\epsilon}\int\!\! \frac{d^{D-1}{\bf q}}{(2\pi)^{D-1}}\frac{e^{i{\bf q}\cdot {\bf x}}-1}{{\bf q}^4}
\right) {\cal G}^{(1)}_{\ref{Fig:vertex:corr:all}b}({\bf p};E) \bigg|_{{\bf q}\;{\rm hard},\;{\bf l}\;{\rm soft}}
\nn\\
&&+\left[\mu^{2\epsilon}Z^2e^4\int\!\! \frac{d^{D-1}{\bf q}}{(2\pi)^{D-1}}\int\!\!
\frac{d^{D-1}{\bf l}}{(2\pi)^{D-1}}\frac{e^{i{\bf q}\cdot {\bf x}}-1}{{\bf q}^2{\bf l}^2({\bf l}-{\bf q})^2}
-\delta Z_{\cal S}\right]
\bigg|_{{\bf q},\:{\bf l}\;{\rm hard}}
\nn\\
& &= c^{(1)}(r)\,{\cal G}^{(1)}_{\ref{Fig:vertex:corr:all}b}({\bf p},E) + c^{(2)}(r)\,{\cal G}^{(0)}({\bf p},E),
\eqa
where the desired Wilson coefficient at order $Z^2\alpha^2$ reads
\bqa
c^{(2)}(\mu,r)\Big|_{\rm MS} &=& \mu^{2\epsilon}Z^2e^4\int\!\! \frac{d^{D-1}{\bf q}}{(2\pi)^{D-1}}\int\!\!
\frac{d^{D-1}{\bf l}}{(2\pi)^{D-1}}\frac{e^{i{\bf q}\cdot {\bf x}}-1}{{\bf q}^2{\bf l}^2({\bf l}-{\bf q})^2}
-\delta Z_{\cal S}
\nn\\
&=& -16\pi^2\mu^{2\epsilon}Z^2\alpha^2  {2^{5-2 D} \pi^{2-\frac{D}{2}}  \over \Gamma \left(\frac{D-2}{2}\right) \sin\frac{\pi  (D-1)}{2}}\int\!\! \frac{d^{D-1}{\bf q}}{(2\pi)^{D-1}}\frac{e^{ i{\bf q}\cdot{\bf r}}-1}{|{\bf q}|^{7-D}}+\frac{Z^2\alpha^2}{2\epsilon}
\nn\\
&=& -Z^2\alpha^2\left(\ln{\mu r}+ {\gamma_E-1+\ln 4\pi\over 2} \right),
\label{c2r:analytical:res:MS:scheme}
\eqa
where the $D$-dimensional expression of $\tilde{c}^{(2)}(q)$ in \eqref{tilde:c2:q} is used.
It is reassuring that $c^{(2)}(r)$ is indeed UV and IR finite, and does contain the $-Z^2\alpha^2 \ln r$ term.
The nonlogarithm constant is scheme-dependent, which is largely irrelevant for our purpose.
Note we have used the simplest MS scheme to renormalize the local operator $\varphi N$.
As indicated in Fig.~\ref{Wilson:coeff:coor:space:decomp}$c)$,
this coefficient receive the contribution solely from the regions
where both loop momenta become hard.
Notice \eqref{Diff:Gamma:10c:and:7e:expand:q:hard}
is compatible with \eqref{Gamma:n:iterative:product:NREFT},
with the new piece of Wilson coefficient given in \eqref{c2r:analytical:res:MS:scheme}.

\subsubsection{Final expression of the OPE in coordinate space,
KG wave function near the origin}

It is the time to sew up everything and represent the OPE relation in coordinate space,
\eqref{OPE:NREFT:coordinate:space} through the order $Z^2\alpha^2$ explicitly as
\beq
\varphi({\bf r})N(0) \xrightarrow{r\to {1\over m}}
\left\{1-mZ\alpha r-Z^2\alpha^2\left(\ln{\mu r}+ {\gamma_E-1+\ln 4\pi\over 2}  \right)
+{\cal O}(Z^3\alpha^3)\right\}[\varphi N]_R(0;\mu)+\cdots.
\label{OPE:coord:space:explicit:formula}
\eeq
This relation holds at operator level, which is the main result of this work.
We stress that both the ${\cal O}(Z^2\alpha^2)$ Wilson coefficient and the renormalized
composite operator $[\varphi N]_R$ become scale-dependent.
If we continue to analyze the asymptotic behaviors of
even higher-order Green functions, it is conceivable that the Wilson coefficients of
the form $Z^3\alpha^3 r \ln r$ and $Z^4\alpha^4 \ln^2 r$ should be correctly extracted.
However, the explicit verification is beyond the scope of the current work,
though we fully trust in the correctness of our approach based on NREFT and OPE.

In \eqref{KG:wf:nonlocal:matrix:element}, we tentatively formulate the KG wave function as the
nonlocal vacuum-to-atom matrix element of the sQED and HNET fields at equal time.
There we fail to use OPE to account for the correct
near-the-origin behavior of the $S$-wave KG wave functions
of hydrogen-like atom.
Encouragingly, the structure of \eqref{OPE:coord:space:explicit:formula} suggests us to
identify the KG wave function of the hydrogen-like atom as the following
equal-time matrix element formed by the product of the NRsQED and HNET fields:
\beq
\Psi_{nlm}({\rm r}) \equiv \langle 0 \vert  \varphi({\bf x}) N(0) \vert n l m \rangle.
\label{KG:wf:nonlocal:matrix:element:NRsQED}
\eeq
When focusing on $S$-wave states, the universal near-the-origin behavior of KG wave
functions in \eqref{KG:WF:origin:Log:Div} and \eqref{KG:Universal:Small:r}, at least to order $Z^2\alpha^2$,
are entirely characterized by the Wilson coefficients associated with the leading $S$-wave
composite operator $[\varphi N]_R$.

Now we come to the crux of this work. It is usually thought that the
Klein-Gordon equation in an external Coulomb field is a genuinely
relativistic wave equation, where one is able to probe arbitrary short-distance
($r\to 0$) profile of wave function.
Nevertheless, one should note that KG equation in \eqref{KG:Equation} is specifically
defined in the rest frame of the nucleus, whose Lorentz covariance is far from obvious.
Actually, our OPE-based analysis reveals that KG wave function should be viewed as a secretely
nonrelativistic wave function, as represented by \eqref{KG:wf:nonlocal:matrix:element:NRsQED} rather than
by \eqref{KG:wf:nonlocal:matrix:element}. As a consequence, it does not make sense to talk about
the near-the-origin behavior of the KG wave function when $r\ll {1\over m}$!
On the contrary, when probing the short-range behavior of the wave function, we are
only justified to let the minimum value of $r$ approach ${1\over m}$.
The reason is simple, since the Compton length of the electron comprises the shortest length scale
where nonrelativistic EFT can still make sense.

\subsection{Renormalization Group Equation for Wilson coefficient}
\label{sec:rge:Wilson:coef}

In the $r\to {1\over m}$ limit, if the renormalization scale $\mu$ is taken around the inverse Bohr radius,
the logarithm $\ln \mu r\approx \ln Z\alpha$ in \eqref{OPE:coord:space:explicit:formula}
is potentially large, which might harm the perturbative convergence of fixed-order prediction.
The convenient and effective method to resum large logarithms to all orders is via
the celebrated renormalization group equation (RGE).
Our starting observation is that the atomic Bethe-Salpeter wave function
in \eqref{KG:wf:nonlocal:matrix:element:NRsQED} is obviously independent of the artificial scale
$\mu$. When $r$ becomes small, one may invoke the OPE relation
\eqref{OPE:NREFT:coordinate:space}, and demand that the combination ${\cal C}(r;\mu) [\varphi N]_R(0;\mu)$
is independent of $\mu$.
The scale dependence of the composite $S$-wave operator
is controlled by the standard RGE:
\beq
\mu {d[\varphi N]_R \over d\mu}=\gamma_S [\varphi N]_R,
\label{S-wave:operator:evolution}
\eeq
with $\gamma_{\mathcal S}=Z^2\alpha^2+{\cal O}(Z^4\alpha^4)$  being the anomalous dimension for the
composite operator $\varphi N$, as given in \eqref{ano-dim}. The solution is
\beq
[\varphi N]_R(\mu) = [\varphi N]_R(\mu_0) \left({\mu\over \mu_0}\right)^{Z^2\alpha^2}.
\eeq

One readily writes down the following RGE for the $S$-wave Wilson coefficient
${\cal C}(r)$:
\beq
\mu {d {\cal C}(r)\over d\mu}+ {\cal C}(r) \gamma_{\mathcal S} = 0.
\label{RGE:cal:C:r:mu}
\eeq
Note $\mu$ and $r$ always come hand in hand inside the logarithm,
dimensional consideration leads to $(\mu {d \over d\mu} - r {d \over dr}){\cal C}(\mu r)=0$.
Define $r\equiv r_0 \kappa$, where $r_0$ is a typical length scale of order Bohr radius.
We can transform the RGE \eqref{RGE:cal:C:r:mu} into
the evolution equation that governs the scaling behavior of ${\cal C}$ near small $r$:
\beq
\kappa {d {\cal C}(r_0 \kappa)\over d\kappa}+ {\cal C}(r_0 \kappa) \gamma_{\mathcal S} = 0.
\label{RGE:cal:C:r:kappa}
\eeq

The solution of \eqref{RGE:cal:C:r:kappa} becomes
\beq
{\cal C}(r) =  {\cal C}(r_0) \kappa^{-Z^2\alpha^2}= {\cal C}(r_0) \left({r\over r_0}\right)^{-Z^2\alpha^2},
\label{RGE:cal:C:r}
\eeq
which has resummed the leading logarithms of $\kappa$ to all orders.
As expected, this anomalous scaling behavior is identical to the near-the-origin behavior of the $S$-wave
KG wave function encoded in \eqref{KG:WF:origin:Log:Div}. Moreover, if setting $\mu=1/r_0= 2/n a_0$
for the $nS$ hydrogen-like atom, we have ${\cal C}(r_0;\mu)=1$. If we further identify the matrix element
$\langle 0\vert [\varphi N]_R(0;\mu=1/r_0)\vert n 00 \rangle$ as the
nonrelativistic Coulomb-Schrodinger wave function at the origin,
we may even fully reproduce \eqref{KG:WF:origin:Log:Div}:
\beq
R^{\rm KG}_{n0}(r) \approx  {\cal C}(r;\mu=1/r_0) \sqrt{4\pi}
\langle 0\vert [\varphi N]_R(0;\mu=1/r_0)\vert n 00 \rangle\approx
\left( {2 r\over n a_0}\right)^{-Z^2\alpha^2} R_{n0}^{\rm Sch}(0).
\label{RGE:OPE:explain:KGWF:near:origin}
\eeq

\section{Summary}
\label{sec:summary}

The Klein-Gordon equation in an external Coulomb field marks an important milestone 
in the early development of relativistic quantum mechanics. It has been known almost a century ago
that the $S$-wave KG wave function for hydrogen-like atoms acquires logarithmic divergence near the origin.
It is often thought that this divergence is of only academic interest, since this logarithmic singularity 
becomes relevant only for extremely small $r$. Moreover, the physical origin underlying this divergent near-the-origin
behavior is rather poorly understood.

As a sequel of our previous work~\cite{Huang:2018yyf}, we employ some modern field-theoretical tools, 
{\it e.g.} EFT and OPE, to solve this long-standing puzzle. We explicitly demonstrate that, by incorporating
the relativistic kinetic correction into the nonrelativistic Coulomb-Schr\"{o}dinger EFT, we are able to
provide a novel perspective to look into this universal logarithmic near-the-origin divergence, {\it i.e.},
which can be interpreted as the Wilson coefficients arising from the product of the nonrelativistic electron field
and HQET-like nucleus field. With the aid of renormalization group equation, we can satisfactorily reproduce the
anomalous scaling behavior observed in the KG wave function at small $r$.  

The folklore says that the Klein-Gordon equation in an external Coulomb field is a  
relativistic wave equation, where one is allowed to probe the arbitrarily small distance  
($r\to 0$) of the wave function. Curiously, our OPE analysis based on relativistic scalar QED 
utterly fails to reproduce the correct short-distance logarithmic divergence of the KG wave function! 
This is a clear sign that the KG equation may secretely describe the nonrelativistic rather than 
relativistic bound-state wave function. In our opinion, it does not make any sense to talk about
the near-the-origin behavior of the KG wave function when $r\ll {1\over m}$, contrary to the folklore.
Furthermore, when probing the short-range behavior of the wave function,  
the minimum value of $r$ is frozen at ${1\over m}$, since the Compton length of the electron 
constitutes the smallest length scale where nonrelativistic EFT can still apply.

In a follow-up work, using the same methodology, we will demonstrate that~\cite{Huang:2019:Dirac},
it is also feasible to employ OPE within nonrelativistic EFT to account for the universal weakly divergent 
near-the-origin behavior of the Dirac wave functions for the $nS_{1/2}$ hydrogen atom.

\begin{acknowledgments}
This work is supported in part by the National Natural Science Foundation of China under Grants No.~11475188,
No.~11621131001 (CRC110 by DFG and NSFC).
\end{acknowledgments}





\begin{thebibliography}{1}

\bibitem{Weinberg:1995mt}
  S.~Weinberg,
  {\it The Quantum theory of fields. Vol.~1: Foundations},
  Cambridge, University Press (1995).

\bibitem{Greiner2000}
W.~Greiner, {\it Relativistic Quantum Mechanics: Wave Equations},
Springer, 2000.



\bibitem{Huang:2019:Dirac}
Y.~Huang, Y.~Jia and R.~Yu, in preparation.

\bibitem{Huang:2018yyf}
  Y.~Huang, Y.~Jia and R.~Yu,
  arXiv:1809.09023 [hep-ph].
  
\bibitem{Wilson:1969zs}
  K.~G.~Wilson,
  Phys.\ Rev.\  {\bf 179}, 1499 (1969).
  
\bibitem{Loewdin1954}
P.-O.~L\"{o}wdin,
Phys.\ Rev.\  {\bf 94}, 1600 (1954).
  
 
\bibitem{Holstein2014}
B.~R. Holstein, {\it Topics in Advanced Quantum Mechanics (Dover Books on
Physics)}, Dover Publications, 2014.
  

\bibitem{Landau:1958}
L.~D.~Landau, E.~M.~Lifshitz, 
{\it Quantum mechanics: non-relativistic theory}, Pergamon press, 1958.

 
\bibitem{Eichten:1989zv}
E.~Eichten and B.~R.~Hill,
Phys.\ Lett.\ B {\bf 234}, 511 (1990).


\bibitem{Georgi:1990um}
H.~Georgi,
	Phys.\ Lett.\ B {\bf 240}, 447 (1990).

\bibitem{Manohar:2000dt}
A.~V.~Manohar and M.~B.~Wise,
{\it Heavy quark physics},
Camb.\ Monogr.\ Part.\ Phys.\ Nucl.\ Phys.\ Cosmol.\  {\bf 10}, 1 (2000).
  
\bibitem{Weinberg:1996kr}
  S.~Weinberg,
{\it The quantum theory of fields. Vol.~2: Modern applications},
Cambridge, University Press (1996).
  
\bibitem{Schwartz:2013pla}
  M.~D.~Schwartz,
{\it Quantum Field Theory and the Standard Model}, Cambridge, 2014.
  
  
\bibitem{Caswell:1985ui}
W.~E.~Caswell and G.~P.~Lepage,
Phys.\ Lett.\  {\bf 167B}, 437 (1986).

\bibitem{Bodwin:1994jh}
  G.~T.~Bodwin, E.~Braaten and G.~P.~Lepage,
  Phys.\ Rev.\ D {\bf 51}, 1125 (1995)
  Erratum: [Phys.\ Rev.\ D {\bf 55}, 5853 (1997)]
  [hep-ph/9407339].

  
  
\bibitem{Brambilla:2004jw}
  N.~Brambilla, A.~Pineda, J.~Soto and A.~Vairo,
  Rev.\ Mod.\ Phys.\  {\bf 77}, 1423 (2005)
  [hep-ph/0410047].
  

\bibitem{Collins:1984xc}
  J.~C.~Collins,
{\it Renormalization : An Introduction to Renormalization, The Renormalization Group, and the Operator Product Expansion},
Cambridge, University Press (1984).


\bibitem{Czarnecki:1997vz}
  A.~Czarnecki and K.~Melnikov,
  Phys.\ Rev.\ Lett.\  {\bf 80}, 2531 (1998)
  [hep-ph/9712222].


\bibitem{Beneke:1997jm}
  M.~Beneke, A.~Signer and V.~A.~Smirnov,
  Phys.\ Rev.\ Lett.\  {\bf 80}, 2535 (1998)
  [hep-ph/9712302].


\bibitem{Luke:1999kz}
  M.~E.~Luke, A.~V.~Manohar and I.~Z.~Rothstein,
  Phys.\ Rev.\ D {\bf 61}, 074025 (2000)
  [hep-ph/9910209].
  



\end{thebibliography}
\end{document}